\documentclass{siamart190516}[review]
\usepackage[utf8]{inputenc}
\usepackage{mathtools}
\DeclarePairedDelimiter\floor{\lfloor}{\rfloor}

\title{A Homogenised Model of Fluid-String Interaction  \thanks{Submitted to the editors
DATE.}
}
\author{ A. Kent \thanks{Mathematical Institute, University of Oxford, Radcliffe Observatory Quarter, Woodstock Road, Oxford OX2 6GG, United Kingdom (kent@maths.ox.ac.uk, waters@maths.ox.ac.uk, oliver@maths.ox.ac.uk, chapman@maths.ox.ac.uk)}
\and S. L. Waters \footnotemark[2]
\and J. Oliver \footnotemark[2]
\and S. J. Chapman \footnotemark[2] }

\begin{document}
\newcommand{\BibTeX}{{\scshape Bib}\TeX\xspace} 
\maketitle
\begin{abstract}
A homogenised model is developed to describe the interaction between aligned strings and an incompressible, viscous, Newtonian fluid. In the case of many strings, the ratio of string separation to domain width gives a small parameter which can be exploited to simplify the problem. Model derivation using multiscale asymptotics results in a modified Darcy law for fluid flow, with coefficients determined by averaged solutions to microscale problems. Fluid flow is coupled to solid deformation via a homogenised force balance obtained by coarse-graining the balance on each string. This approach offers an alternative method to systematically derive the equations governing the interaction of Stokes flow with many flexible structures. The resulting model of fluid-structure interaction is reduced to a single scalar, linear, partial differential equation by introducing a potential for the pressure. Analytical solutions are presented for a cylindrical geometry subject to time harmonic motion of the string ends. Scaling laws are identified that describe the variation of shear stress exerted on the string surface with the forcing frequency.
\end{abstract}


\section{Introduction}
Studies of the interaction between thin flexible structures and slow Newtonian flow have been motivated by applications ranging from biological sensing \cite{Lauga2016}, tissue engineering and microtubule dynamics \cite{Nazockdast2017} through to textile production \cite{Terrill1994}. 
The model we develop is motivated by a particular tendon tissue engineering setup, where cells are grown on bundles of many aligned strings held under tension \cite{Mouthuy2022}. The strings can be deformed, inducing a flow to provide mechanical stimulation to the growing cells which aims to mimic physiological stresses. Tendon cells modulate their behaviour in response to mechanical stresses, motivating the development of models which can capture the coupling between macroscale forcing and microscale shear stress \cite{Maganaris2017}. 

Several mathematical and numerical techniques have been developed to simulate the interaction between single flexible fibres and Stokesian flow, reviewed in \cite{DuRoure2019, Duprat2016}. Whilst an effective way to describe the behaviour of a small number of fibres, these methods can become unwieldy for many-fibre systems. 

Insights into multi-fibre systems have progressed through the development of theoretical descriptions for `hairy' versions of canonical experiments in fluid dynamics \cite{Hosoi2019}. Enhanced entrainment of a viscous fluid during dip-coating of a fibre-covered plate has been described theoretically by treating the gaps between rigid fibres as individual channels \cite{Nasto2018, Nasto2016}.  The models developed in \cite{Nasto2018, Nasto2016} are applicable in a regime where viscous effects dominate elasticity. When the bending stiffness of the fibres is smaller, viscous and elastic effects balance. Hydrodynamic stresses can induce solid deformation which in turn alters the distribution of obstacles experienced by the flow. Such two-way coupling between fluid and solid deformation was captured in \cite{Alvarado2017}, where drag laws characterising Couette flow over a fibre bed were established, including flow rectification for angled fibres. Fibres were modelled as nonlinear beams assumed to deform identically, with fluid stress modelled as a point force at the beam tip. In some situations, the fluid-structure interaction induces spatially varying deformations in the solid phase meaning that the assumption of identical deformation breaks down. 

Capturing spatially varying fluid-structure interaction can be achieved by coarse-grained models which describe the solid deformation and fluid motion using continuum fields \cite{Gopinath2011,Stein2019, Szaniawski1977}. We now outline coarse-grained continuum models and methods of derivation in view of their application to modelling the interaction between Stokes flow and many-fibre systems.

Volume averaging is one method of continuum model derivation, where the governing equations are averaged over a representative elementary volume (REV) which captures the microscale properties of a material \cite{Drew1983}. The resulting averaged equations describe materials constructed from many REVs. Volume averaging has been applied to model the deformation of sparse fibre beds in response to slow Newtonian flow, by defining separate averages along the fibre direction and in the fluid phase \cite{Stein2019}. Coupling the fibre deformation to the background flow field using local slender body theory, the averaged model takes the form of a Brinkman-like equation for the fluid velocity relative to that of the fibre bed \cite{Brinkman1949, Stein2019}. In the dilute limit, the hydrodynamic interactions between fibres are weak, allowing models to be constructed by summing over contributions to fluid flow from the interaction with each individual structure \cite{ Stein2019, Thomazo2020}.

At higher solid volume fractions it is necessary to account for the hydrodynamic interactions between fibres, as the flow disturbance induced by the motion of one fibre influences the deformation of its neighbours. At such volume fractions, viscous dissipation predominantly occurs on the scale of separation between obstacles, and the flow is described by a generalised Darcy's law \cite{Gopinath2011}. Poroelastic equations with Darcy-like flow have been used to describe disordered polymer beds, ordered fibre beds and the viscous washing of fibre bundles in textile production \cite{Gopinath2011, Ockendon1993, Szaniawski1977, Terrill1994}. For poroelastic materials, continuum momentum and mass conservation equations are generally written for the solid and fluid phases, coupled by the pore pressure. Poroelastic equations describing fibrous media can be obtained by coarse-graining continuum equations for a fluid and elastic solid \cite{Gopinath2011}. 
Assuming \textit{a priori} a continuum, linear elastic model for the solid phase introduces parameters such as the shear modulus which are typically determined experimentally \cite{Gopinath2011}. Beginning with a discrete model for each individual fibre surrounded by a fluid continuum and homogenising, a macroscale model can be derived for the fluid-fibre interaction where the coefficients are determined by the averaged solution of an independent microscale problem \cite{Artini2017, Artini2018}. 

Homogenisation via multiscale asymptotics is a coarse-graining technique which exploits the regularity of a material on a length scale much smaller than the whole structure (the microscale) to derive the effective properties of the material \cite{Hinch1991}.  Typically the domain is assumed to be exactly periodic, constructed from translations of a single `unit cell'. The microscale, also known as the fast scale, describes variations over a unit cell, while the macroscale, or slow scale, describes variations over the whole domain. Homogenisation of the Stokes equations describing fluid flowing through a solid with low porosity results in Darcy's law \cite{Holmes2013, Keller1980, Levy1983}. If the solid phase is assumed to be an elastic continuum, the equations of poroelasticity are obtained after homogenisation \cite{Burridge1981}.  Instead of two interpenetrating continua, the problem can be formulated by considering a force balance on discrete solid elements surrounded by fluid. This approach was taken by Artini and Gianluca \cite{Artini2017, Artini2018} to obtain homogenised equations for the interaction of rigid cylinders tethered to their initial positions by springs with surrounding transverse flow governed by the unsteady Stokes equations. 

To fully justify the derivation of homogenised equations from force balances on discrete solid elements, several extensions to the standard method of multiple scales are required.
Balancing fluid traction with a restoring force results in an integral equation at the microscale - the integral of the fluid stress around the string boundary appears in the force balance. Examples of problems featuring integral constraints include wave propagation through bubbly liquids, calculating the electric potential within nematic crystals, and radiative transfer in porous media \cite{Bennett2018, Chapman2015, Rooney2021}. Integral constraints need to be handled very carefully when using the method of multiple scales. For point-wise boundary conditions, the macroscale coordinate can be treated as constant within a given unit cell. As illustrated in \cite{Chapman2015, Rooney2021}, this approach fails for integral constraints due to terms arising from the small changes in the macroscale coordinate at different points in a given unit cell. 

In considering string deformations that vary across the domain, the assumption of exact periodicity of the microscale geometry must be relaxed. Slow scale variation in the microscale problem can be captured, normally at the cost of solving the microscale problem multiple times for various values of the slow scale position which parameterise the problem \cite{Bruna2015, Dalwadi2015, Richardson2011}. However, the interplay between these two extensions - integral constraints and a slowly varying unit cell - is nontrivial, and introduces further complications.

In this paper, we present a continuum model for the interaction between aligned strings and Stokes flow using homogenisation via multiscale asymptotics, systematically incorporating the extensions to standard multiple scales outlined above. A force balance on individual strings gives rise to an integral condition, with variations in displacement across the domain creating a slowly varying microstructure. Transverse deformations of the strings are considered, coupled to the surrounding fluid flow. This methodology complements existing coarse-grained models, offering a means to model systems with a higher solid volume fraction than some volume averaging techniques \cite{Stein2019}, while maintaining some description of the features of individual solid elements which can be lost when posing solid continuum models \textit{a priori} \cite{Gopinath2011}. 

The  model setup is outlined in section \ref{sec:setup} and nondimensionalised in section \ref{sec:nondim}. The equations are homogenised in section \ref{sec:homog}, where we obtain three microscale problems and one macroscale problem describing the interaction of fluid and strings as an effective medium. Solutions to the microscale and macroscale problems are presented in sections \ref{sec:cellsolutions} and \ref{sec:macsolutions} respectively. Conclusions are drawn in section \ref{sec:discussion}.

\section{Model Setup}\label{sec:setup}
\begin{figure}
    \centering
    \includegraphics[width = 0.9\textwidth]{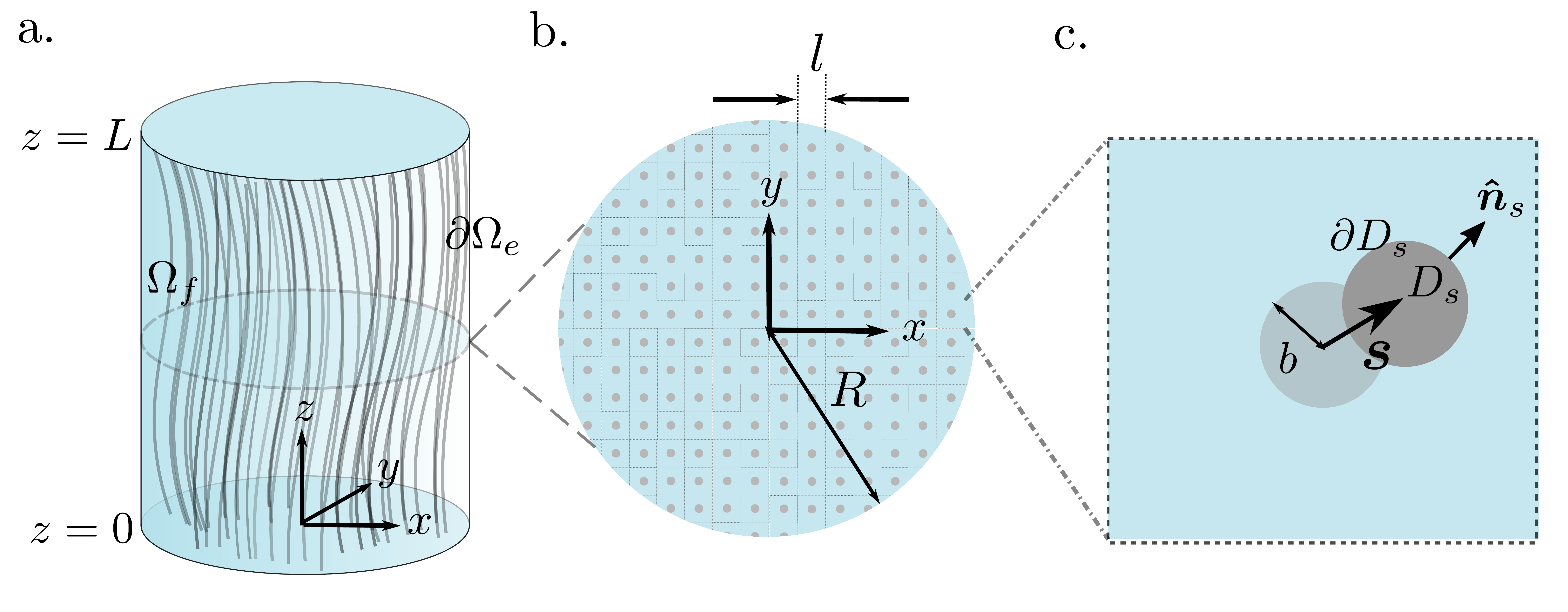}
    \caption{Schematic of the model setup. a. Viscous, incompressible, Newtonian fluid in the region $\Omega_f$ surrounds strings aligned under tension in an examplar cylindrical domain with radius $R$ and length $L$. Fluid is represented in blue and strings are drawn in grey. b. The approximate representation of the domain cross-section used in the homogenisation procedure. c. `Zooming' into the macroscale domain, indicated by the dot-dashed lines, shows the unit cell which forms the domain for the microscale problem. Dashed lines show the boundaries where periodicity is imposed.}
    \label{fig:modelschem}
\end{figure}

We consider an incompressible, Newtonian fluid with dynamic viscosity $\mu$, surrounding $N$ aligned strings separated by a distance $l$, each with circular cross-section of radius $b$. The superscript $i$ indexes the string under consideration: $i = 1, 2, ...,N$. We use Cartesian coordinates, taking the origin of the $z$-axis to be at the string base, with the $z$-axis aligned with the strings' equilibrium configuration.  The initial position of the centre-line of the $i$th string $\boldsymbol{x}_0^i$ is assumed to lie on a square periodic lattice which covers the whole domain, as illustrated in figure \ref{fig:modelschem}.b. The curved exterior boundary of the fluid domain is $\partial \Omega_e$ and the flat boundary is denoted $\Omega_{\perp}$. The domain has characteristic dimension $L$, with a fluid region $\Omega_f$. The region occupied by a string in a transverse cross-section is $D_s^i$ and the boundary of the cross-section is $\partial D_s^i$. String motion imposed at one end drives fluid flow within the domain. 

We model the fibres as uniform, elastic strings under tension executing small, transverse displacements. Assuming the string cross-section does not deform, we can describe the displacement of the $i$th string from its reference configuration via the displacement of its centre-line $\boldsymbol{s}^i(z, t) = (s_1^i, s_2^i)$.

The surrounding flow is modelled as a slow, incompressible, Newtonian fluid governed by
\begin{align}
    -\mu \nabla^2 \boldsymbol{u} + \nabla p &= \boldsymbol{0},\label{eq:stokes} \\
    \nabla \cdotp \boldsymbol{u} &= 0 \label{eq:incomp} .
\end{align}
where $\boldsymbol{u} = u \boldsymbol{\hat{e}}_x + v \boldsymbol{\hat{e}}_y + w\boldsymbol{\hat{e}}_z$ is the fluid velocity and  $p$ the fluid pressure. For convenience, we decompose the velocity into its transverse and axial components, introducing $\boldsymbol{v} = (u, v)$. 

No-slip conditions are imposed at the boundary of each string so that
\begin{align}
    \boldsymbol{v} &= \frac{\partial \boldsymbol{s}^i}{\partial t} \quad \text{on} \quad \partial D_s^{i} \label{eq:bcnosliprod}, \\
    w &= 0 \quad \text{on} \quad \partial D_s^i, \label{eq:bcnsrodw}
\end{align}
where $t$ denotes time, for each $0 \leq z \leq L.$
We also require boundary conditions for the fluid flow on the outer edges of the domain. The exact form that these boundary conditions take will be given in section \ref{sec:homprob} after derivation of the homogenised model. Homogenisation reduces the order of the fluid problem and thus reduces the number of boundary conditions required.

We balance the fluid force with elastic forces due to the string tension. The nondimensional number describing the relative strength of the string inertia to the string tension is $\rho T b^2 / \mu^2 L^2 $ for a string with mass volume density $\rho$. We consider strings with an aspect ratio $b/L$ sufficiently small that $\rho T b^2 / \mu^2 L^2 \ll 1$.
Neglecting the inertial terms, the transverse force balance on the $i$th string is given by
 \begin{equation}
     \int_{\partial D_s^{i}} \boldsymbol{\sigma}_{\perp} \cdotp \boldsymbol{\hat{n}}_s dl + \frac{\partial}{\partial z} \left( T \frac{\partial \boldsymbol{s}^i}{\partial z}   \right) = \boldsymbol{0},\label{eq:forcebalance}
 \end{equation}
 where $T$ is the tension, $\boldsymbol{\hat{n}}_s$ is the outward pointing unit normal to the string surface and $\sigma_{\perp}$ is the $2\times 2$ fluid stress tensor for the transverse flow components. The axial force balance is given by
 \begin{equation}
     \int_{\partial D_s^i} \boldsymbol{\sigma}_\parallel \cdotp \boldsymbol{\hat{n}}_s dl + \frac{\partial T}{\partial z} = 0, \label{eq:fbax}
 \end{equation}
where $\boldsymbol{\sigma}_{\parallel} = \mu(u_z + w_x, v_z + w_y)$.
Despite neglecting inertia in both the fluid and solid momentum balances, we require an initial condition on the displacement which we set as 
\begin{equation}
    \boldsymbol{s}^i = \boldsymbol{0}.\label{eq:incs}
\end{equation}
The strings are fixed at their base and a displacement is imposed at the upper end,
\begin{align}
      \boldsymbol{s}^i &= \boldsymbol{0} \quad \text{at} \quad z = 0,\\
      \boldsymbol{s}^i &= \boldsymbol{a}^i(t) \quad  \text{at} \quad z =  L\label{eq:bcstringtop}
\end{align}
Various examples of imposed in-plane displacements will be considered in section \ref{sec:macsolutions}.

\section{Nondimensionalisation and Scaling}\label{sec:nondim}
We assume that the characteristic lengthscale of separation between adjacent fibres is much less than width of the domain, and introduce the small parameter $ \delta = l/L \ll 1 \label{eq:delta}$. 
Motivated by the scalings used to derive Darcy's law from homogenisation of the Stokes equations on a porous domain, we seek a balance between macroscopic pressure gradients and microscale viscous terms. We nondimensionalise the problem by scaling
 \begin{equation}
 \begin{split}
      (u, v, w) = U(\hat{u}, \hat{v}, \hat{w}), &\quad  (x, y, z) = L(\hat{x}, \hat{y}, \hat{z}),  \\
     \boldsymbol{s}^i = \delta L \boldsymbol{\hat{s}}^i, \quad \boldsymbol{a}^i = \delta L \boldsymbol{\hat{a}}^i, & \quad p = p_a + \frac{\mu U }{\delta^2 L} \hat{p},  \quad t = \delta \frac{L}{U}\hat{t}, \label{eq:scales}
\end{split}
 \end{equation}
 where $p_a$ is the atmospheric pressure and hats denote nondimensional quantitities.
 The velocity scale will be determined later by choosing scalings to balance terms in the string force balance. The timescale is chosen to ensure we have a balance at leading-order in (\ref{eq:bcnosliprod}) so that string displacement drives fluid motion. This balance results in the presence of string velocity in the leading-order macroscale equations. 
 
 After nondimensionalisation and dropping hats, Stokes equations (\ref{eq:stokes})-(\ref{eq:incomp}) become
 \begin{align}
     -\delta^2 \nabla^2 \boldsymbol{u} + \nabla p &=0, \label{eq:nondimstokes}\\
     \nabla \cdotp \boldsymbol{u} &= 0 \label{eq:nondimincomp}.
 \end{align}
 No-slip conditions on the string boundaries (\ref{eq:bcnosliprod})-(\ref{eq:bcnsrodw}), gives
 \begin{align}
     \boldsymbol{v} &= \frac{\partial \boldsymbol{s}^i}{\partial t} \quad \text{on} \quad \partial D_s^i,  \label{eq:nondimbcnsrod} \\
     w &= 0 \quad \text{on} \quad \partial D_s^i. \label{eq:bcnsw}
 \end{align}
 Nondimensionalisation of the force balance (\ref{eq:forcebalance}) yields
\begin{equation}\label{eq:nondimforcebalance}
    \int_{\partial D_s^i} \boldsymbol{\sigma}_{\perp} \cdotp \boldsymbol{\hat{n}}_s  dl  +
    T^* \frac{\partial }{\partial z}\left( \mathcal{T} \frac{\partial \boldsymbol{s}^i}{\partial z} \right)
    = \boldsymbol{0}.
\end{equation}
The nondimensional parameter $T^*$ is the ratio of the tension to viscous terms, given by
 \begin{equation}
     T^* = \frac{T \delta}{\mu U L}
 \end{equation}
 and $\mathcal{T} = \mathcal{T}(z)$ is a nondimensional $O(1)$ function describing axial variations in tension. 
 We set $T^* = 1$, fixing the velocity scale of the problem as
 \begin{equation}
     \quad U = \frac{\delta T }{\mu L}.
 \end{equation}
 After nondimensionalising the axial force balance (\ref{eq:fbax}), we obtain
 \begin{equation}
       \delta \int_{\partial D_s^i} \boldsymbol{\sigma}_\parallel \cdotp \boldsymbol{\hat{n}}_s dl + \frac{\partial \mathcal{T}}{\partial z} = 0.
 \end{equation}
 At leading-order, we see the tension is uniform along the string length.  Henceforth, we consider $\mathcal{T} = 1.$ At higher orders, we expect that the differences in fluid shear along the string length will induce changes in tension, though the first order correction to the tension does not feature in our analysis. 
 Nondimensionalising the boundary conditions on string displacement gives
 \begin{align}
      \boldsymbol{s}^i&= \boldsymbol{0} \quad \text{at} \quad z = 0,\\
      \boldsymbol{s}^i &= \boldsymbol{a}^i(t)  \quad \text{at} \quad z = 1,
 \end{align}
and the initial condition (\ref{eq:incs}) becomes
\begin{equation}
    \boldsymbol{s}^i= \boldsymbol{0} \quad \text{at} \quad t =0. 
\end{equation}

\section{Homogenisation}\label{sec:homog}
In this section, we use multiscale asymptotics to derive continuum equations describing the fluid and string motion on the macroscale. Instead of solving coupled equations for the string motion and flow field in a complicated domain, the aim is to obtain two simpler problems that can be solved separately: one determining microscale variations in the flow field around a single string, the second effective medium equations for the fluid-string interaction. 

The slow scale $\boldsymbol{x}$ describes variation on the scale of the domain width, i.e. the macroscale, and we introduce a fast scale $\boldsymbol{X} = (x, y) /\delta$ which describes variation on the scale of the string separation, i.e. the microscale. We assume that these two scales are independent, and that the domain cross-section can be constructed from unit cells that each contain a single fibre, as shown in figure \ref{fig:modelschem}.c. We assume that the number of strings is sufficiently large that boundary effects from the breakdown of periodicity at the edge of the domain are negligible.

\subsection{Continuum Equations for String Displacements}\label{sec:scont}
We begin by writing equations describing the string displacements in a form that can be homogenised. We assume that the displacement of neighbouring strings is sufficiently similar to allow the introduction of a continuum field for the string displacement. 
We introduce piece-wise constant functions for the string displacement and initial position
\begin{align}
    \boldsymbol{s}(\boldsymbol{x}, t) &=  \boldsymbol{s}^i(z, t) \quad  \text{in} \quad  D^i_s,\label{eq:pwconst} \\
    \boldsymbol{x}_0(\boldsymbol{x}) &= \boldsymbol{x}^i_0(z) \quad \text{in} \quad  D^i_s
\end{align}
defined over the solid region of the unit cell $D^i_s$. The function $\boldsymbol{s}$ takes on the value of the displacement of the string found in the same unit cell as $\boldsymbol{x}$. Equation (\ref{eq:pwconst}) cannot be transformed into multiple scales form easily and so we follow \cite{Chapman2015} in rephrasing this equation as
\begin{equation}
    \nabla_{\perp} \boldsymbol{s} = 0 \quad \text{in} \quad D_s, \label{eq:string}
\end{equation}
where $\nabla_{\perp} = (\partial_x, \partial_y).$

\subsection{Multiple Scales form of the Governing Equations}
To write the differential equations in multiple scales form, we apply the chain rule to the 2D gradient operator 
\begin{equation}
    \nabla_{\perp}  = \nabla_x + \frac{1}{\delta}\nabla_X,\label{eq:ms}
\end{equation}
where $\nabla_x = (\partial_x, \partial_y)$ and $\nabla_X  = (\partial_X, \partial_Y)$. We have introduced the fast variable only in the transverse plane, leaving the $z$-coordinate untouched.
We seek multiple scales expansions for the fluid velocity, pressure and string displacement by expanding as a power series in $\delta$
\begin{align}
    \boldsymbol{v} &= \boldsymbol{v}_0(\boldsymbol{x}, \boldsymbol{X}, t) + \delta \boldsymbol{v}_1(\boldsymbol{x}, \boldsymbol{X}, t) + O(\delta^2) \label{eq:ums},\\
    w  &= w_0(\boldsymbol{x}, \boldsymbol{X}, t) + \delta w_1(\boldsymbol{x}, \boldsymbol{X}, t) + O(\delta^2) \label{eq:wms},\\
    p  &= p_0(\boldsymbol{x}, \boldsymbol{X}, t) + \delta p_1(\boldsymbol{x}, \boldsymbol{X}, t) + O(\delta^2) \label{eq:pms},\\
    \boldsymbol{s} &= 
   \boldsymbol{s}_0(\boldsymbol{x}, \boldsymbol{X}, t) + \delta \boldsymbol{s}_1(\boldsymbol{x}, \boldsymbol{X}, t) + O(\delta^2)\label{eq:sms},
\end{align}
where the variables in all expansions are periodic in $\boldsymbol{X}$ with period $\boldsymbol{1}$, motivated by the microscale structure of the domain. We substitute the expansions (\ref{eq:ums})-(\ref{eq:sms}) into the Stokes equations (\ref{eq:nondimstokes})-(\ref{eq:nondimincomp}) and the equation governing string displacement (\ref{eq:string}). 
At leading-order in $\delta$, we find
\begin{align}
    \nabla_X p_0 &= \boldsymbol{0} \quad \text{in} \quad D_f \label{eq:o1p},\\
    \nabla_X \cdotp \boldsymbol{v}_{0} & = 0 \quad \text{in} \quad D_f  \label{eq:o1incomp},\\
    \nabla_{X} \boldsymbol{s}_0 &= 0  \quad \text{in} \quad D_s \label{eq:o1s}.
\end{align}
Equation (\ref{eq:o1p}) implies that the leading-order pressure $p_0 = p_0(\boldsymbol{x}, t)$ and string displacement $\boldsymbol{s}_0 = \boldsymbol{s}_0 (\boldsymbol{x}, t)$ are independent of the fast scale. At $O(\delta)$, 
\begin{align}
   - \nabla^2_X \boldsymbol{v}_0 + \nabla_x p_0 + \nabla_X p_1 &= \boldsymbol{0} \quad \text{in} \quad D_f,  \label{eq:odeltap}\\
   - \nabla^2_X w_0 + \frac{\partial p_0}{\partial z} &= 0  \quad \text{in} \quad D_f,  \label{eq:odeltaw} \\
    \nabla_X \cdotp \boldsymbol{v}_{ 1} + \nabla_x \cdotp \boldsymbol{v}_{0} + \frac{\partial w_0}{\partial z}&= 0 \quad \text{in} \quad D_f , \label{eq:odeltaincomp}\\
    \nabla_{x} \boldsymbol{s}_0 + \nabla_{X} \boldsymbol{s}_1 &= 0. \quad \text{in} \quad D_s . \label{eq:odeltas}
\end{align}
We can integrate (\ref{eq:odeltas}) with respect to the fast variable to find
\begin{equation}
    \boldsymbol{s}_1 = - \boldsymbol{X} \cdotp \nabla_x \boldsymbol{s}_0 + \boldsymbol{c}_0,\label{eq:s1}
\end{equation}
where $\boldsymbol{c}_0 = \boldsymbol{c}_0(\boldsymbol{x})$ is a function from integration.
Before deriving the resulting microscale and macroscale problems, we write the boundary conditions (\ref{eq:nondimbcnsrod})-(\ref{eq:nondimforcebalance}) in multiple scales form.

\subsection{Transforming Boundary Conditions into Multiple Scales Form}\label{sec:hombc}
Homogenisation via multiscale asymptotics in multiple spatial dimensions generally relies on the assumption of periodicity in the microscale variables in order for the microscale and macroscale equations to decouple. Progress can also be made in scenarios where the unit cell is slowly varying \cite{Bruna2015, Dalwadi2015, Richardson2011}. At present, our microscale problem is solved over the domain shown in figure \ref{fig:modelschem}.c. The position of the string boundary in this domain $\partial D_s $ is given by the level set $h_s = 0$ of the function
\begin{equation}
    h_s = |\boldsymbol{X} - \boldsymbol{X}_0 - \boldsymbol{s}| - B, \label{eq:lss}
\end{equation}
where $B = b/\delta$ is the string radius and $\boldsymbol{X}_0 = \boldsymbol{x}_0/ \delta - \floor{\boldsymbol{x}_0/\delta}$ is the initial position in the $(x, y)$-plane measured with respect to the microscale coordinates. Thus, the string boundary is dependent on both the fast and slow scales through the string displacement (\ref{eq:sms}). Substituting (\ref{eq:sms}) into (\ref{eq:lss}), we find the leading-order string boundary $\partial D_{s_0}$ is defined to be the level set $h = 0$ of the function
\begin{equation}
    h = |\boldsymbol{X} - \boldsymbol{X}_0  - \boldsymbol{s}_0 | - B. \label{eq:lolevelset}
\end{equation}
As the leading-order string displacement is independent of the fast scale, the microscale domain with this boundary will be locally periodic. This results in macroscale and microscale problems that decouple, facilitating the homogenisation process. We begin by expanding the no-slip condition at the string boundary onto the leading-order domain before transforming the force balance into multiple scales form.

\subsubsection{No-slip Conditions}\label{sec:noslip}
We expand the no-slip condition at the string boundary (\ref{eq:nondimbcnsrod}) about the leading-order position $\partial D_{s_0}$ in the fast variable. As the leading-order position depends on the slow scale, we have an $O(\delta)$ contribution to the no-slip condition from the variation in the leading-order displacement as we move around the boundary. This variation arises due to the small changes in the value of $\boldsymbol{x}$ as we move around the unit cell. In principle, this variation should be taken into account when evaluating the boundary conditions in all multiple scales problems. However for most problems the resulting terms cancel (being slow derivatives of a lower order balance), so they are almost always ignored, consciously or unconsciously. In the present problem, where the string position is an unknown function of position, one of these terms remains, so we need to treat the boundary conditions carefully. 

We can account for this variation explicitly using the approach developed in \cite{Chapman2015} for integral constraints. The slow-fast interdependence can be dealt with by considering deviations of a macroscale point from a fixed reference point. That is, we write $\boldsymbol{x} = \boldsymbol{\hat{x}} - \delta \boldsymbol{b} + \delta \boldsymbol{X}$, where $\boldsymbol{\hat{x}}$ is an arbitrary reference point in the same unit cell as $\boldsymbol{x}$, and $\boldsymbol{b} = \boldsymbol{\hat{x}}/\delta - \lfloor \boldsymbol{\hat{x} } / \delta \rfloor$ is a constant vector joining the $\boldsymbol{\hat{x}}$ to the left-hand corner of the unit cell. Henceforth, we take our reference point $\boldsymbol{\hat{x}}$ to be the lower left corner of the unit cell, fixing $\boldsymbol{b}  = \boldsymbol{0}.$ Thus if $\boldsymbol{x}$ is a point on the boundary of the string, then the velocity $\boldsymbol{u}$ is evaluated not at $(\boldsymbol{x}, \boldsymbol{X})$, but rather at $(\boldsymbol{\hat{x}}+ \delta \boldsymbol{X}, \boldsymbol{X})$.
 
We expand about the leading-order position, for example
\begin{align} \label{eq:bcexp}
    \boldsymbol{v}_0(\boldsymbol{\hat{x}} + \delta \boldsymbol{X}, \boldsymbol{X}^b  +\boldsymbol{s}_0(\boldsymbol{\hat{x}} &+ \delta \boldsymbol{X}) +  \delta \boldsymbol{s}_1 (\boldsymbol{\hat{x}} + \delta \boldsymbol{X} ,  \boldsymbol{X}  )  + ... , t) = \\ 
    & \boldsymbol{v}_0 + \delta \boldsymbol{X} \cdotp \nabla_x \boldsymbol{v}_0 + \delta (\boldsymbol{s}_1 + \boldsymbol{X}\cdotp \nabla_x \boldsymbol{s}_0) \cdotp \nabla_X \boldsymbol{v}_0 + O(\delta^2)\nonumber ,
\end{align}
where $\boldsymbol{X}^b$ are points on the initial position of the string boundary and terms in the expansions are evaluated at $\boldsymbol{x} = \boldsymbol{\hat{x}}, \boldsymbol{X} = \boldsymbol{X}^b + \boldsymbol{s}_0(\boldsymbol{\hat{x}})$. Taking a similar expansion for all terms in the no-slip boundary condition, we find
\begin{align}
 \boldsymbol{v}_0 +  \delta \boldsymbol{X} \cdotp \nabla_x \boldsymbol{v}_0 + & \delta (\boldsymbol{s}_1 + \boldsymbol{X}\cdotp \nabla_x \boldsymbol{s}_0) \cdotp \nabla_X \boldsymbol{v}_0  + \delta \boldsymbol{v}_1  = \label{eq:noslipexp}\\ &\frac{\partial \boldsymbol{s}_0}{\partial t}  +\delta \boldsymbol{X}\cdotp \nabla_x  \left(\frac{\partial \boldsymbol{s}_0}{\partial t}  \right) + \delta  \frac{\partial \boldsymbol{s}_1}{\partial t} + O(\delta^2) \quad \text{on} \quad \partial D_{s_0}, \nonumber 
\end{align}
where terms involving the fast gradient of the leading-order string velocity vanish due to (\ref{eq:o1s}). 
 Comparing coefficients at each order of $\delta$, we obtain the following conditions on the leading-order fluid velocity
\begin{equation}
    \boldsymbol{v}_0 = \frac{\partial \boldsymbol{s}_0}{\partial t} \quad \text{on} \quad \partial D_{s_0} .\label{eq:bco1ns}
\end{equation}
Using (\ref{eq:bco1ns}), we can simplify (\ref{eq:noslipexp}) by cancelling the slow gradient of the leading-order velocity terms that appear on both sides of the equation at $O(\delta)$. The condition on the first order fluid velocity becomes
\begin{equation}
 \boldsymbol{v}_1 = \frac{\partial \boldsymbol{s}_1}{\partial t} - (\boldsymbol{s}_1 +  \boldsymbol{X} \cdotp \nabla_x \boldsymbol{s}_0) \cdotp \nabla_X \boldsymbol{v}_0  \quad \text{on} \quad \partial D_{s_0}.\label{eq:odeltans}
\end{equation}

We can write (\ref{eq:odeltans}) in a more convenient form using the solution to the $O(\delta)$ string problem.
Substituting the expression for $\boldsymbol{s}_1$ into (\ref{eq:odeltans}), we find
\begin{equation}
 \boldsymbol{v}_1 = \frac{\partial \boldsymbol{s}_1}{\partial t} - \boldsymbol{c}_0 \cdotp \nabla_X \boldsymbol{v}_0 \quad \text{on} \quad \partial D_{s_0} \label{eq:bcodns}.
\end{equation}
In a naive treatment of (\ref{eq:bcnosliprod}), the term involving $\boldsymbol{c}_0$ would not be present, highlighting the need for a careful treatment of the boundary. Repeating the analysis on (\ref{eq:bcnsw}), we obtain
\begin{equation}
    w_0 = w_1 =  0  \quad \text{on} \quad \partial D_{s_0}.
\end{equation}

\subsubsection{Force Balance}\label{sec:integralfb}
To transform the force balance on the string into multiple scales form, we need to expand the integral of fluid traction on the string about the leading-order domain; account for the variation of the slow scale with the fast scale as we move around the string boundary; and expand the integrand in multiple scales form.

A detailed justification of this procedure, which extends the analysis in \cite{Chapman2015} to the case of slowly varying unit cells, is given in \cite{us2022}. In essence, we calculate the size of the the contribution from the variation in the value of the slow scale around the string boundary in a similar way to section \ref{sec:noslip}, by writing a point on the string boundary $\boldsymbol{x}$ in terms of its microscale distance from a fixed macroscale point $\boldsymbol{\hat{x}}$. Following \cite{us2022}, and using the expansion (\ref{eq:bcexp}) to expand this slow-fast interdependence onto the leading-order boundary position, we find
\begin{equation}
\begin{split}
       \int_{\partial D_{s}} \boldsymbol{\sigma}_{\perp} \cdotp \boldsymbol{\hat{n}}dl  = \delta \int_{\partial D_{s_0}} & \boldsymbol{\sigma}_{\perp}  \cdotp \boldsymbol{\hat{n}} dl_X +   \delta^2 \int_{\partial D_{s_0}} ( \boldsymbol{s}_1 + \boldsymbol{X} \cdotp \nabla_x \boldsymbol{s}_0) \cdotp \nabla_X  \boldsymbol{\sigma}_{\perp}  \cdotp \boldsymbol{\hat{n}} dl_X \\
       & +\delta^2 \int_{\partial D_{s_0}}  \boldsymbol{X} \cdotp \nabla_x \boldsymbol{\sigma}_{\perp}  \cdotp \boldsymbol{\hat{n}} dl_X + O(\delta^3), \label{eq:intexp0}
\end{split}
\end{equation}
where $dl_X$ is the line element in fast coordinates and $\boldsymbol{\hat{n}}$ is the outward-facing normal to the boundary of the leading order string position. 
Substituting the following multiple scales expansion for the fluid stress, 
\begin{equation}
    \boldsymbol{\sigma}_{\perp}  = -\delta^{-2}p_0 I_2 + \delta^{-1} [-p_1I_2 + \nabla_X \boldsymbol{v}_0 + (\nabla_X \boldsymbol{v}_0)^T] + O(\delta^2),
\end{equation}
where $I_2 = \rm{diag}(1, 1)$, into (\ref{eq:intexp0}) gives
\begin{equation}
       \int_{\partial D_{s}} \boldsymbol{\sigma}_{\perp} \cdotp \boldsymbol{\hat{n}}dl  =   \int_{\partial D_{s_0}} [-p_1I_2 + \nabla_X \boldsymbol{v}_0 + (\nabla_X \boldsymbol{v}_0)^T]\cdotp \boldsymbol{\hat{n}} dl_X  - \int_{\partial D_{s_0}} \boldsymbol{X} \cdotp \nabla_x p_0 \cdotp \boldsymbol{\hat{n}} dl_X +O(\delta)
\label{eq:intexp}
\end{equation}
where we have used (\ref{eq:o1p}). Applying the divergence theorem to each term and periodicity, we find
\begin{equation}
        \int_{\partial D_{s}} \boldsymbol{\sigma}_{\perp} \cdotp \boldsymbol{\hat{n}}dl  =  - \int_{D_{f}} \nabla_X \cdotp [-p_1I_2 + \nabla_X \boldsymbol{v}_0 + (\nabla_X \boldsymbol{v}_0)^T] dS_X  - \int_{ D_{s_0}}  \nabla_{x} p_0  dS_X +O(\delta),\label{eq:intintexp}
\end{equation}
where $dS_X = dX dY$ denotes the area element in fast scale variables. Since $p_0$ is independent of the fast scale (see \ref{eq:o1p}), it is straightforward to integrate $\nabla_x p_0$ over the string domain $D_{s_0}.$ Using (\ref{eq:odeltap}) we can evaluate both terms in (\ref{eq:intintexp}) to obtain
\begin{equation}
      \int_{\partial D_{s}} \boldsymbol{\sigma}_{\perp} \cdotp \boldsymbol{\hat{n}}dl  =   - \nabla_{x} p_0  + O(\delta)\label{eq:int}.
\end{equation}
Had we not accounted for the variation in the slow scale across the unit cell the right-hand side of (\ref{eq:int}) would have been $-\phi_f \nabla_{x} p_0$, so that the forcing from fluid shear would have been underestimated by a factor of $\phi_s \nabla_{x} p_0$. Here $\phi_s = \pi B^2$ denotes the solid volume fraction and $\phi_f = 1 - \phi_s$ the fluid volume fraction. We deduce from (\ref{eq:nondimforcebalance}) and (\ref{eq:int}) that the leading-order force balance at the string boundary takes the form
\begin{equation}
    - \nabla_{x} p_0 +   \frac{\partial^2 \boldsymbol{s}_0}{\partial z^2} = \boldsymbol{0}.\label{eq:bcfbms}
\end{equation}
The homogenised force balance (\ref{eq:bcfbms}) forms part of the
macroscale problem. This will be coupled to equations describing macroscale variation of the fluid. We return to the homogenisation of the fluid problem, beginning with the derivation of the cell problems. 

\subsection{Cell Problems}\label{sec:microprob}
Cell problems are solved to determine the microscale behaviour of the dependent variables, and their averaged solutions provide the coefficients for the macroscale problem. 
We obtain the cell problems from equation (\ref{eq:o1incomp}), (\ref{eq:odeltap}) and (\ref{eq:odeltaw}) which we restate here for convenience:
\begin{align}
   \nabla_X \cdotp \boldsymbol{v}_{0}&= 0 \quad \text{in} \quad D_f, \label{eq:cellincomp}\\
   - \nabla^2_X \boldsymbol{v}_0 + \nabla_x p_0 + \nabla_X p_1 &= \boldsymbol{0} \quad \text{in} \quad D_f, \label{eq:cellmomt}\\
   - \nabla^2_X w_0 + \frac{\partial p_0}{\partial z} &= 0  \quad \text{in} \quad D_f,  \label{eq:cellmoma}
\end{align}
with no-slip on the string boundary 
\begin{align}
    \boldsymbol{v}_0 &= \frac{\partial \boldsymbol{s}_0 }{\partial t}  \quad \text{on} \quad \partial D_{s_0}, \\
    w_0 &= 0 \quad \text{on} \quad \partial D_{s_0} \label{bc:cellnsw},
\end{align}
and periodicity of $ \boldsymbol{v}_0, w_0, p_0$ and $p_1$ in $\boldsymbol{X}$ with period $\boldsymbol{X}_p$.
Exploiting the slow scale independence of the leading-order pressure and string velocity, we decompose our solution for fluid velocity and pressure as follows
\begin{align}
    \boldsymbol{v}_{0} &= \Psi_{\perp} \cdotp \nabla_x p_0 +  \Pi \cdotp \frac{\partial \boldsymbol{s}_0}{\partial t} \label{eq:microu},\\
    w_0 &= \Psi_{33} \frac{\partial p_0}{\partial z}\label{eq:microw},\\
    p_1 &= \boldsymbol{\Phi}\cdotp \nabla_x p_0 + \boldsymbol{\xi } \cdotp \frac{\partial \boldsymbol{s}_0}{\partial t} + c_1,
\end{align}
where $c_1 = c_1(\boldsymbol{x}, t)$ is a function independent of the fast scale determined by the solvability condition for the microscale problem at next order. The functions $(\Psi_\perp, \boldsymbol{\Phi})$ (a $2\times 2$ tensor and vector respectively) satisfy the inhomogeneous problem derived from (\ref{eq:cellmomt}) and (\ref{eq:cellincomp}) with homogeneous boundary conditions, given by
\begin{align}
   -  \nabla_X^2 \Psi_\perp +  \nabla_X \boldsymbol{\Phi} + I &= 0 \quad \text{in}\quad D_f   \label{eq:psimom} , \\
    \nabla_X \cdotp \Psi_\perp &= 0 \quad \text{in}\quad D_f, \\
    \Psi_\perp \text{ and } \Phi \text{ periodic and } \Psi_{\perp} &= 0 \quad \text{on} \quad \partial D_{s_0}\label{eq:psibc}.
\end{align}
The cell functions $(\Pi, \boldsymbol{\xi})$ satisfy the homogeneous problem with inhomogeneous boundary conditions i.e.
\begin{align}
         - \nabla_X^2 \Pi + \nabla_X \boldsymbol{\xi} &= 0 \quad \text{in}\quad D_f   \label{eq:pimom},\\
    \nabla_X \cdotp \Pi &= 0  \quad \text{in}\quad D_f ,  \\
     \Pi \text{ and }  \xi  \text{ periodic and } \Pi &= I \quad \text{on} \quad \partial D_{s_0} \label{eq:pibc}.
\end{align}
The microscale variation of the axial flow is given by (\ref{eq:cellmoma}), (\ref{bc:cellnsw}) and (\ref{eq:microw}) 
\begin{align}
    \nabla^2_X \Psi_{33} &= 1 \quad \text{in} \quad D_f, \label{eq:wcell} \\
    \Psi_{33} &= 0 \quad \text{on} \quad \partial D_{s_0}, \label{eq:bcwcell}
\end{align}
with $\Psi_{33}$ is periodic in $\boldsymbol{X} $.

\begin{figure}
    \centering
    \includegraphics[width = 0.8\textwidth]{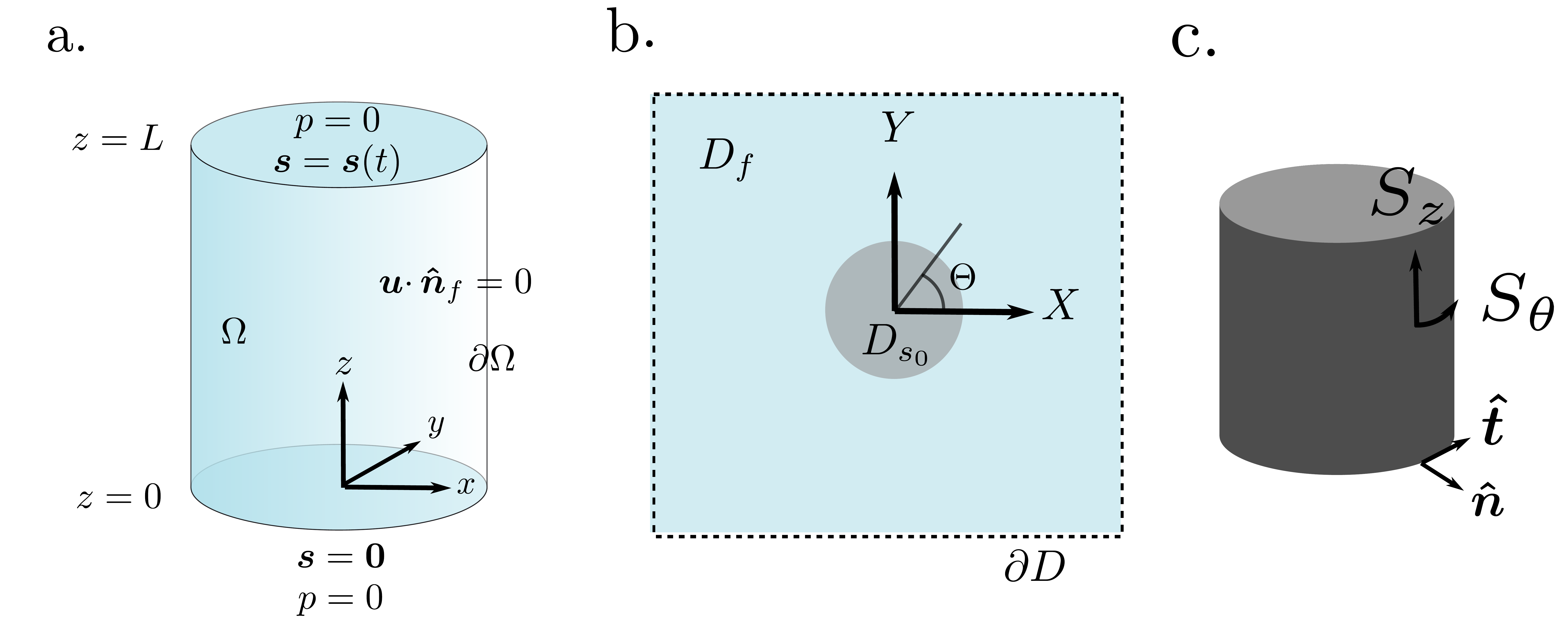}
    \caption{After homogenisation the problem is reduced to two simpler problems. a. The domain of the macroscale problem where the fluid and strings are treated as a continuum. b. Domain of the microscale problem, whose averaged solutions give permeability coefficient in the macroscale equations. c. Directions of fluid-induced shear stress at the string surface.}
    \label{fig:domain}
\end{figure}

The microscale problems are defined over the domain of the unit cell, as illustrated in figure \ref{fig:domain}.b.  The unit cell domain defined by (\ref{eq:lolevelset}) will contain a single string located at a position set by the macroscale coordinate of the unit cell - its displacement from the exactly periodic reference configuration is given by $\boldsymbol{s}_0 = \boldsymbol{s}_0(\boldsymbol{x}, t)$.  Using the periodicity of boundary conditions, we can equivalently solve over the domain $D = D_f \cup D_{s_0}$ (shown in figure  \ref{fig:domain}.b) where the string is placed at the origin of microscale coordinates.

The problem driven by inhomogeneous boundary conditions has an analytical solution,
\begin{equation}
    \Pi = I \quad \text{and} \quad \boldsymbol{\xi} = \boldsymbol{0}.\label{eq:pisol}
\end{equation}
We will present numerical solutions to the cell problems (\ref{eq:psimom})-(\ref{eq:psibc}) and (\ref{eq:wcell})-(\ref{eq:bcwcell}) in section \ref{sec:cellsolutions}. As the fast scale $\boldsymbol{X}$ has been introduced in the transverse plane, the problems are solved over a 2-dimensional domain. This will give rise to an anisotropic permeability as we have different problems determining the axial and transverse components of the permeability tensor. The effects of solid fraction on anisotropy will be explored in section \ref{sec:cellsolutions}. 

\subsection{Homogenised Problem}\label{sec:homprob}
The averaged solutions to the microscale problems (\ref{eq:psimom})-(\ref{eq:psibc}), (\ref{eq:pimom})-(\ref{eq:pibc}) and (\ref{eq:wcell})-(\ref{eq:bcwcell}) provide the coefficients in the homogenised problem which describes variations in fluid flow and string deformation across the domain. To derive the macroscale equations, we integrate the incompressibility condition at first order (\ref{eq:odeltaincomp}) across the fluid domain of the unit cell $D_f$, obtaining
\begin{equation}
    \int_{D_f} \nabla_X \cdotp \boldsymbol{v}_{ 1} dS_X + \int_{D_f} \nabla_x \cdotp \boldsymbol{v}_{0}dS_X + \int_{D_f} \frac{\partial w_0}{\partial z} dS_X = 0.
\end{equation}
We apply the divergence theorem to the first integral and Reynolds' transport theorem to the final two terms to give
\begin{equation}
\begin{split}
    - \int_{\partial D_{s_0}} \boldsymbol{v}_{1 } \cdotp \boldsymbol{\hat{n}} dl_X + \nabla_x \cdotp \int_{D_f} \boldsymbol{v}_{0} dS_X  + \int_{\partial D_{s_0}} \boldsymbol{v}_{0  } \cdotp \nabla_x \boldsymbol{s}_0 \cdotp \boldsymbol{\hat{n}} dl_X \\
    + \frac{\partial}{\partial z} \int_{D_f} w_0 dS_X + \int_{\partial D_{s_0}}w_0 \frac{\partial \boldsymbol{s}_0}{\partial z} \cdotp \boldsymbol{\hat{n}}dl_X   = 0 \label{eq:intmid},
\end{split}
\end{equation}
where the periodicity of $\boldsymbol{v}_1$ means that its integral over the outer boundary of the unit cell, $\partial D$, vanishes. The boundary terms from applying the Reynolds transport theorem evaluate to zero as the axial component vanishes on the boundary and the integrand involving the transverse velocity is independent of the fast scale; (\ref{eq:bco1ns}) implies that $\boldsymbol{v}_0$ is independent of the fast scale on the boundary. Applying the no-slip boundary condition (\ref{eq:bcodns}) to the first term of (\ref{eq:intmid}), we find
\begin{equation}
    - \int_{\partial D_{s_0}}\left(  \frac{\partial \boldsymbol{s}_1}{\partial t} - \boldsymbol{c}_0 \cdotp \nabla_X \boldsymbol{v}_{0 }  \right)\cdotp \boldsymbol{\hat{n}} dl_X + \nabla_x \cdotp \int_{D_f} \boldsymbol{v}_{0 } dS_X + \frac{\partial}{\partial z} \int_{D_f} w_0 dS_X = 0. \label{eq:homder}
\end{equation}
We use the divergence theorem to take the first term into the solid domain and use the trace of (\ref{eq:odeltas}), which gives
\begin{equation}
\begin{split}
     - \int_{\partial D_{s_0}}\frac{\partial \boldsymbol{s}_1 }{\partial t} \cdotp \boldsymbol{\hat{n}} dl_X  = - \int_{D_{s_0}} \nabla_X \cdotp \left(  \frac{\partial \boldsymbol{s}_1}{\partial t}  \right) & dS_X  = \int_{D_{s_0}} \nabla_x \cdotp \left(\frac{\partial \boldsymbol{s}_0}{\partial t} \right) dS_X \\ &= \phi_s \nabla_x \cdotp \left( \frac{\partial \boldsymbol{s}_0}{\partial t}\right)\label{eq:w1},
\end{split}
\end{equation}
where we have used (\ref{eq:o1s}) in evaluating the final integral in (\ref{eq:w1}).
We apply the divergence to the second term of (\ref{eq:homder}), taking the integral into the fluid region to give
\begin{equation}
      \int_{\partial D_{s_0}} c_{0i}  \frac{\partial v_{0j}}{\partial X_i} \hat{n}_{0j} dl_X = \int_{D_f} c_{0i} \frac{\partial^2 v_{0j}}{\partial X_j\partial X_i} dS_X = 0,
\end{equation}
where we have used the fast scale independence of $\boldsymbol{c}_0$, periodicity of $\boldsymbol{u}_0$ and (\ref{eq:o1incomp}).  

Thus, (\ref{eq:homder}) reduces to 
\begin{equation}
    \nabla_x \cdotp \langle \boldsymbol{v}_{0  } \rangle    + \frac{\partial \langle w_0 \rangle}{\partial z}= -\frac{\phi_s}{\phi_f} \nabla_{x} \cdotp  \left( \frac{\partial \boldsymbol{s}_0}{\partial t} \right) \label{eq:homfluida},
\end{equation}
where angled brackets denote the spatial average over the fluid domain in the unit cell, defined as 
\begin{equation}
    \langle g \rangle = \frac{1}{\phi_f} \int_{D_f} g dS_X. 
\end{equation}
Writing $\boldsymbol{u} = \langle \boldsymbol{u}_0 \rangle$, $\nabla = (\partial_x, \partial_y, \partial_z)$, and dropping subscripts on leading-order quantities henceforth, (\ref{eq:homfluida}) becomes
\begin{equation}
    \nabla \cdotp \boldsymbol{u} = - \frac{\phi_s}{\phi_f} \nabla_{\perp} \cdotp \left( \frac{\partial \boldsymbol{s}}{\partial t} \right) \label{eq:homfluid}
\end{equation}
where $\nabla_{\perp} = (\partial_x, \partial_y)$.
Averaging (\ref{eq:microu})-(\ref{eq:microw}) gives
 \begin{align}
   \boldsymbol{v}&= - \mathcal{K}_{\perp} \nabla_{\perp} p + \frac{\partial \boldsymbol{s}}{\partial t},\label{eq:homu} \\
   w &= -\kappa_3 \frac{\partial p}{\partial z} \label{eq:homw}
 \end{align} 
where 
 \begin{equation}
     \mathcal{K}_{\perp} = - \langle \Psi_{\perp} \rangle  \quad \text{ and } \quad \kappa_3 =  - \langle  \Psi_{33} \rangle \label{eq:average}
 \end{equation}
are given by the averaged solutions to the cell problems (\ref{eq:psimom})-(\ref{eq:psibc}) and (\ref{eq:wcell})-(\ref{eq:bcwcell}). In evaluating $\langle \Pi \rangle = 1$, we have used (\ref{eq:pisol}).

 The coefficients $\mathcal{K}$ and $\langle \Pi \rangle$ give the contribution to fluid flow from macroscopic pressure gradients and string drag respectively. 
  Combined with the force balance (\ref{eq:bcfbms}), 
\begin{equation}
    - \nabla_{\perp} p + \frac{\partial^2 \boldsymbol{s}}{\partial z^2}  = \boldsymbol{0},\label{eq:homfb}
\end{equation}
we have a system of equations (\ref{eq:homfluid}), (\ref{eq:homu}), (\ref{eq:homfb}) describing the coupling between string displacement and fluid motion. We find the relative fluid velocity satisfies Darcy's law with the condition for mass conservation including a source term due to the presence of strings. Here, equation (\ref{eq:homfluid}) reflects the influx of fluid into regions of the domain where strings move apart and vice versa. We can eliminate fluid velocity from the homogenised equations, substituting (\ref{eq:homu})-(\ref{eq:homw}) into (\ref{eq:homfluid}) to obtain the following
 \begin{align}
     \nabla \cdotp \left( \mathcal{K} \nabla p \right) &= \frac{1}{\phi_f} \nabla_{\perp} \cdotp \left( \frac{\partial \boldsymbol{s}}{\partial t} \right)\label{eq:masscon} ,
 \end{align}
 which is coupled to (\ref{eq:homfb}). We have introduced the permeability tensor which has entries
 \begin{equation}
    \mathcal{K} = - \begin{pmatrix}
    \langle \Psi_{\perp_{11}} \rangle && \langle \Psi_{\perp_{12}} \rangle && 0 \\
   \langle  \Psi_{\perp_{21}} \rangle && \langle \Psi_{\perp_{22}} \rangle && 0 \\
   0 && 0&& \langle \Psi_{33} \rangle  \\
    \end{pmatrix},
\end{equation}
combining the solutions to the cell problems for transverse and axial flow. 
 Here, the mass conservation condition (\ref{eq:masscon}) is the same as that derived for linearly elastic media saturated by viscous fluid in \cite{Barry2001, Gopinath2011}. The momentum equation (\ref{eq:homfb}) agrees with the model consisting of continuum constituents developed in \cite{Gopinath2011}, where the stress of the composite is constructed from a linear superposition of macroscale pressure gradients and elastic stress.

 The accompanying boundary conditions are those of flux or pressure for the fluid and string displacement at each end. We are not able to impose no-slip on the outer edge of the domain as the homogenisation process has  resulted in a lower order model for the fluid flow. We impose no flux of fluid through the outer boundary
 \begin{equation}
     \boldsymbol{u} \cdotp \boldsymbol{\hat{n}}= 0\quad \text{on} \quad \partial \Omega_e,
 \end{equation}
 and a prescribed pressure at the top and base of the domain,
  \begin{equation}
    p = 0 \quad \text{at}\quad z = 0, 1\label{eq:bcp}.
 \end{equation}
We assume that the strings are pinned in their undeformed position at the base
 \begin{equation}
     \boldsymbol{s} = \boldsymbol{0} \quad \text{at} \quad z = 0
 \end{equation} 
and the motion of the upper end of the strings is prescribed
\begin{equation}
    \boldsymbol{s} = \boldsymbol{a}(x, y, t) \quad \text{at} \quad z = 1.\label{eq:bcstop}
\end{equation}

\section{Solution}\label{sec:solutions}
In this section, we describe the methods used to solve the cell problems (\ref{eq:psimom})-(\ref{eq:psibc}) and (\ref{eq:wcell})-(\ref{eq:bcwcell}) and the homogenised problem (\ref{eq:homfluid}), (\ref{eq:homu}) and (\ref{eq:homfb}).

\subsection{Cell Problem Solution}\label{sec:cellsolutions}
The solutions to the cell problems provide insight into the microscale variation of fluid variables and, when averaged, provide coefficients in the macroscale equations. 
The microscale problems (\ref{eq:psimom})-(\ref{eq:psibc}) and (\ref{eq:wcell})-(\ref{eq:bcwcell}) describing how pressure gradients induce fluid flow were solved using the finite element method with FEniCS \cite{AlnaesBlechta2015}. Taylor-Hood elements were used to solve (\ref{eq:psimom})-(\ref{eq:psibc}) and first order Lagrange elements for (\ref{eq:wcell})-(\ref{eq:bcwcell}). 

\begin{figure}
    \centering
    \includegraphics[width = \textwidth]{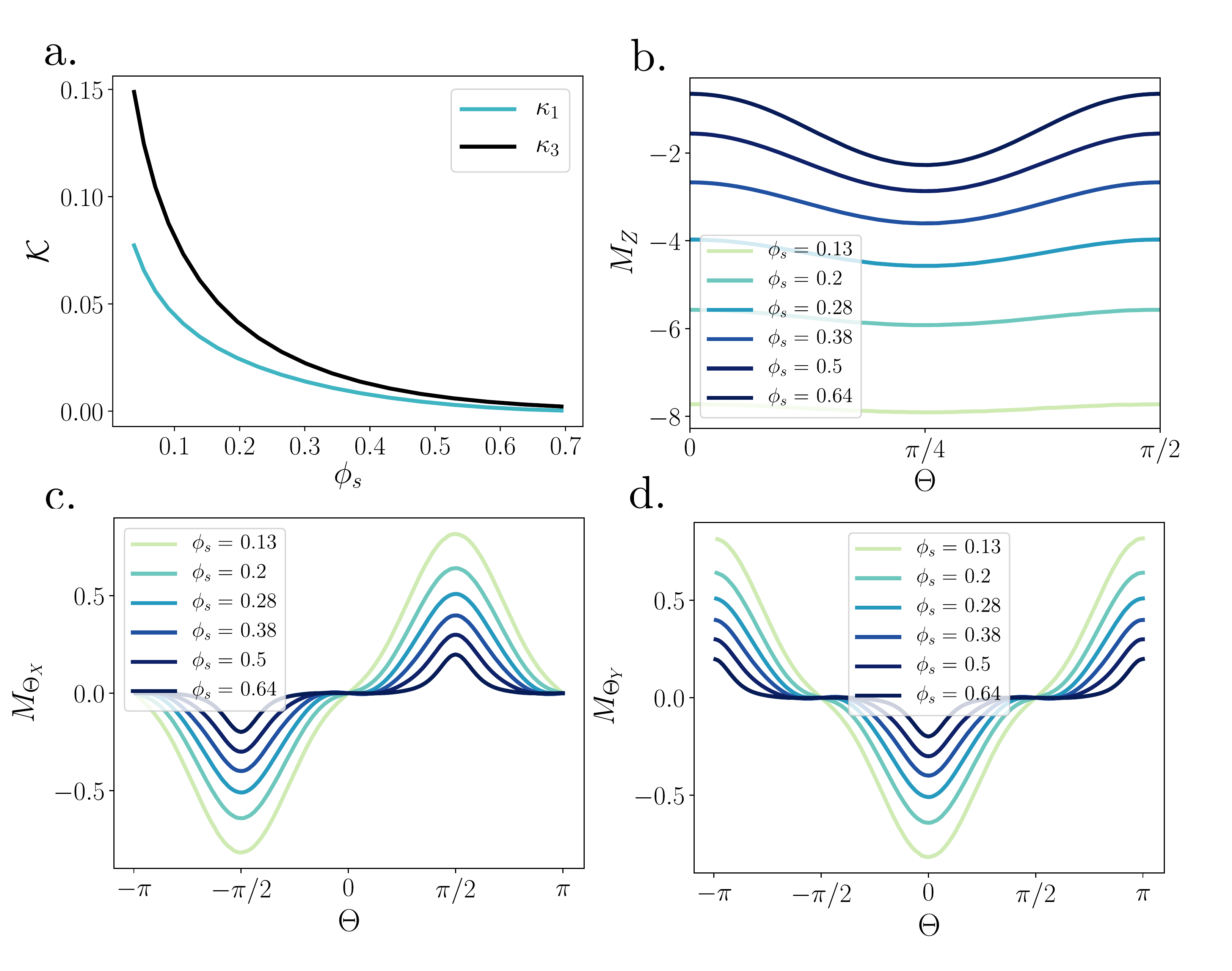}
    \caption{a.) Averaged solutions to the microscale problem give the coefficients which appear in the macroscale problem. We solve the microscale problem for strings with circular cross-section on the domain given in figure \ref{fig:domain}. (b.) Microscale variations in axial shear stress. Symmetry implies the variation shown repeats in all four quadrants.  (c.)-(d.) Microscale variations in the transverse components of shear stress.}
    \label{fig:micro_shear}
\end{figure}

While we will require the solution for all elements in (\ref{eq:psimom})-(\ref{eq:psibc}) to calculate the microscale shear stress, considering symmetry in the specific case of strings with a circular cross-section allows us to deduce the form of the permeability tensor. 
The transformation $ X \to -X $ and $Y \to -Y$ leaves the boundary conditions, domain and forcing invariant, so we expect the solution of (\ref{eq:psimom})-(\ref{eq:psibc}) to be invariant under this transformation. This holds providing the components of $\Psi_{\perp}$ satisfy
\begin{align}
    \Psi_{\perp_{12}}(X, Y) &= \Psi_{\perp_{12}}(-X, Y), \quad  \Psi_{\perp_{12}}(X, Y) = -\Psi_{\perp_{12}}(X, -Y), \\
      \Psi_{\perp_{21}}(X, Y) &= -\Psi_{\perp_{21}}(-X, Y), \quad  \Psi_{\perp_{21}}(X, Y) = \Psi_{\perp_{12}}(X, -Y).
\end{align}
Given the definition of the average (\ref{eq:average}), the contributions of $\Psi$ to the off-diagonal components of the permeability tensor vanish so that the permeability tensor takes the form $\mathcal{K} =\rm{diag}(\kappa_1, \kappa_1, \kappa_3) $. The permeability coefficients for different solid fractions are plotted in figure \ref{fig:micro_shear}.a. As expected, the permeability decreases as the solid fraction increases. With the permeability of an array of cylinders being inversely proportional to the fluid drag on a given cylinder, we have a lower drag from strings aligned with the flow direction and thus have $\kappa_3 > \kappa_1.$ At low solid volume fractions, the axial permeability is twice that of the transverse permeability. Estimates for the permeability of arrays of rigid cylinders have been calculated using various methods, see for example \cite{Drummond1984, Happel1959, Sparrow1959}. Analytical methods approximating the low density limit with flow around a single cylinder recover the ratio of 2 between the axial and transverse permeability \cite{Drummond1984}. 

In addition to providing coefficients in the averaged equations, solutions to the cell problems give the microscale variation in shear stress. The fluid-induced traction exerted on the string surface is given by $\tau \boldsymbol{\hat{n}}$, where $\tau$ is the deviatoric stress tensor which has nondimensional form $\boldsymbol{\tau} = \nabla  \boldsymbol{u} + (\nabla \boldsymbol{u})^T$, and $\boldsymbol{\hat{n}}$ is the outward-facing normal to the leading-order position of the string boundary. 
After nondimensionalisation with respect to the scales given in (\ref{eq:scales}), and transforming to multiple scales form using (\ref{eq:ms}), we find the leading-order axial and transverse shear stress are
\begin{equation}
    S_Z = \frac{1}{\delta}\boldsymbol{\hat{n}} \cdotp \nabla_{X} w 
\quad \text{and} \quad
    S_{\Theta} = \frac{1}{\delta} \left[\sin(2 \Theta ) (v_Y - u_X ) +  \cos(2 \Theta) (u_Y + v_X)  \right]\label{eq:shear}
\end{equation}
respectively, where $\Theta$ is the angle measured from the $X$-axis in microscale coordinates (figure \ref{fig:domain}). 
We calculate the gradient of the velocity with respect to the microscale variables using the expression for velocity given in the derivation of the cell problems (\ref{eq:microu}) and (\ref{eq:microw}).
As the microscale problem for $\Pi$ has a constant solution, there is no contribution to microscale shear stress from string motion. Instead, the strings contributes to the microscale variations via their impact on the domain geometry, while the magnitude of the shear stress is modulated by macroscale pressure gradients. We introduce the functions $M_Z$ and $\boldsymbol{M}_{\Theta}$ which capture the microscale variation in shear stress
\begin{align}
  M_Z &= \frac{1}{\delta} \boldsymbol{\hat{n}} \cdotp \nabla_{X}\Psi_{33}\\
  M_{\Theta_k} &= \frac{1}{\delta} \hat{t}_i \left( \partial_{X_i} \Psi_{\perp_{jk}} + \partial_{X_j} \Psi_{\perp_{ik}}  \right) \hat{n}_j, \quad k = 1,2.
\end{align}
where $\boldsymbol{\hat{t}}$ is the unit tangent to the string boundary in the transverse plane.
The axial and transverse shear stress are then given by 
\begin{equation}
    S_Z  = M_Z p_z \quad \text{and} \quad 
    S_{\Theta} = \boldsymbol{M}_{\Theta} \cdotp \nabla_{\perp} p
\end{equation}
respectively. 
Variations in $M_Z$ and $\boldsymbol{M}_{\Theta}$ at different points on the string boundary are plotted in figure \ref{fig:micro_shear}. The components of $\boldsymbol{M}_{\Theta}$ increase in magnitude with decreasing solid volume fraction, reflecting the higher velocity gradients established when imposing a fixed pressure drop across a larger cross-sectional area. 

\subsection{Macroscale Problem Solution}\label{sec:macsolutions}
To characterise the fluid flow and string deformation across the whole domain, we solve the macroscale equations (\ref{eq:homfb})-(\ref{eq:masscon}) with parameters set by the averaged solution to microscale problems as described above in section \ref{sec:cellsolutions}. In section \ref{sec:potential}, we show that the problem can be simplified by defining a potential related to the fluid pressure. We consider flow driven by string motion in sections \ref{sec:platemotion} and \ref{sec:radial}. 

\subsubsection{Potential Formulation}\label{sec:potential}
To facilitate analysis, we introduce a fluid potential $\phi$ such that $p = \varphi_{zz}$. This allows us to reduce (\ref{eq:homfb})-(\ref{eq:masscon}) to a single, scalar partial differential equation that can be solved analytically in simple geometries. The corresponding displacement is obtained by integrating the homogenised force balance (\ref{eq:homfb}) twice with respect to $z$,
\begin{equation}\label{eq:pots}
    \boldsymbol{s} = \nabla_{\perp} \varphi + \begin{pmatrix} \partial_y \psi_1 \\ - \partial_x \psi_1   \end{pmatrix}z +  \begin{pmatrix} \partial_y \psi_2 \\ - \partial_x \psi_2   \end{pmatrix}.
\end{equation}
We have absorbed the components of the vector functions of integration that may be written as the gradient of a scalar function into the definition of $\varphi$. The functions $\psi_i = \psi_i(x, y)$ for $i = 1, 2$ are the remaining functions of integration. Equation (\ref{eq:masscon}) then gives
\begin{equation}
    \nabla \cdotp \left( \mathcal{K} \nabla \varphi_{zz} \right) = \frac{1}{\phi_f} \nabla_{\perp}^2 \varphi_t  \label{eq:macroprob}.
\end{equation}
Imposing string displacement at the top and base of the domain leads to a problem for $\psi_i = \psi_i(x, y)$. We can eliminate $\varphi$ by cross-differentiating the boundary condition at the base which results in a problem for $\psi_1$. Repeating the process at $z =1 $ gives a problem for $\psi_2$. After solving, we are left with a single partial differential equation for $\varphi.$ We will see examples of this process in sections \ref{sec:platemotion} and \ref{sec:radial}.

The remaining boundary conditions on the fluid potential $\varphi$, correspond to zero imposed pressure 
\begin{equation}
    \varphi_{zz} = 0
\end{equation}
or no fluid flux 
\begin{equation}
    \left( -\mathcal{K} \nabla \varphi_{zz} + \nabla_{\perp_3} \varphi_t + \nabla \times (\psi_{1t} \boldsymbol{\hat{e}}_z ) z  + \nabla \times (\psi_{2t} \boldsymbol{\hat{e}}_z)  \right)\cdotp  \boldsymbol{\hat{n}} =0 ,\label{eq:bcphiflux}
\end{equation}
obtained by substituting the potentials into the expression for fluid velocity (\ref{eq:homu})-(\ref{eq:homw}) in the flux condition $\boldsymbol{u} \cdotp \boldsymbol{\hat{n}} = q_i$. We use $\nabla_{\perp_3} = (\partial_x, \partial_y, 0)$ to denote the in plane gradient.
 The fluid potential $\varphi$ is defined up to linear function of $z$ with constant coefficients, and the displacement potentials $\psi_i$ are defined up to an arbitrary constant. We are free to choose a Dirichlet boundary condition on $\psi_1, \psi_2$ on $\partial \Omega_e$, after which they are determined uniquely. We choose $\psi_1 = \psi_2 = 0$ on $\partial \Omega_e$.  We must also impose an initial condition on the potential which can be obtained from the initial displacement field (\ref{eq:incs}). 
 In the next section, we solve the problem for different boundary conditions.

\subsubsection{Rigid Body Motion of an Upper Plate}\label{sec:platemotion}
In this section, we consider a cylindrical domain illustrated in figure \ref{fig:domain}. The cylinder has nondimensional radius $R = 1/2$ and length $L = 1$. We take the origin of the coordinates to be at the centre of the base, with $z$-axis aligned with the cylinder's axis. We will use both Cartesian coordinates and cylindrical polars $(r, \theta, z)$ to describe the problem, denoting the polar unit vectors $\boldsymbol{\hat{e}}_r, \boldsymbol{\hat{e}}_\theta$ and $\boldsymbol{\hat{e}}_z.$
Supposing strings are fixed to a rigid, porous plate at the top and bottom of the domain, we consider the effects of imposing an in-plane rigid body motion on the top plate. We assume there is no vertical pressure drop across the domain and that no fluid flows through the curved outer boundary of the cylinder. 

We consider an imposed time-harmonic motion of the upper string end, decomposed into a translation and rotation, as follows
\begin{equation}
    \boldsymbol{s}  = \begin{pmatrix}
    A_1 \\
    A_2 
    \end{pmatrix} \cos(\omega t)+ A_3 \begin{pmatrix}
     y\\
     -x  
    \end{pmatrix}\cos(\omega t) \quad \text{at}\quad z = 1, \label{eq:bcsz1}
\end{equation}
where $\omega$ is the frequency of oscillation, nondimensionalised on the timescale given in (\ref{eq:scales}) as $\hat{\omega} = \mu L^2 \omega/T.$
The position of the strings at the base of the domain is fixed (figure \ref{fig:domain}.a),
\begin{equation}
    \boldsymbol{s} = \boldsymbol{0} \quad \text{at}\quad z = 0. \label{eq:bcsz0}
\end{equation}
There is no pressure drop across the domain 
\begin{equation}
    \varphi_{zz} = 0 \quad \text{on} \quad z = 0, 1,\label{eq:bcphizzz}
\end{equation}
and no flux of fluid through the outer wall of the domain $\partial \Omega_e$
\begin{equation}
       \left( -\mathcal{K} \nabla \varphi_{zz} + \nabla_{\perp_3} \varphi_t + \nabla \times (\psi_{1t} \boldsymbol{\hat{e}}_z ) z  + \nabla \times (\psi_{2t} \boldsymbol{\hat{e}}_z)  \right)\cdotp  \boldsymbol{\hat{n}} = 0, \quad \text{on} \quad \partial \Omega_e.\label{eq:bcplateflux}
\end{equation}

We look to solve (\ref{eq:macroprob}) subject to (\ref{eq:bcsz1})-(\ref{eq:bcplateflux}) by seeking time harmonic solutions
\begin{equation}
      \varphi =   \Re[G(x, y, z) e^{i \omega t} ] , \quad 
    \psi_1 = \Re[H_1(x, y, z) e^{i \omega t }], \quad 
    \psi_2 = \Re[H_2(x, y, z) e^{i \omega t} ], \label{eq:thsol}
\end{equation}
which reduces (\ref{eq:macroprob}) to
\begin{equation}
    \nabla \cdotp \left( \mathcal{K} \nabla G_{zz} \right)  = \frac{i \omega}{\phi_f} \nabla^2_{\perp}G \label{eq:macrog}.
\end{equation}
We now look to convert the boundary conditions (\ref{eq:bcsz1})-(\ref{eq:bcplateflux}) into conditions that determine $H_1$ and $H_2$. Following this, we can identify boundary conditions for $G$ to close (\ref{eq:macrog}). 

Beginning by considering the boundary condition on string displacements (\ref{eq:bcsz1})-(\ref{eq:bcsz0}), we can establish a problem for $H_i$ via (\ref{eq:pots}). Written in terms of potentials, the boundary condition (\ref{eq:bcsz0}) on displacement at $z = 0$ becomes
\begin{align}
   \partial_x G +  \partial_y H_2 &= 0 \label{eq:bcspz0} ,\\
   \partial_y G - \partial_x H_2 &= 0 \label{eq:bcspz1}.
\end{align}
We cross-differentiate and subtract (\ref{eq:bcspz0}) from (\ref{eq:bcspz1}) to find
\begin{equation}
    \nabla_{\perp}^2 H_2 = 0 \quad \text{in} \quad \Omega_{\perp},
\end{equation}
where $\Omega_{\perp}$ is the transverse cross-section of $\Omega$ at $z = 0$.
Since we chose to impose
\begin{equation}
    H_2 = 0 \quad \text{on} \quad \partial \Omega_{\perp},
\end{equation}
we find $H_2 = 0$. Undertaking a similar process with (\ref{eq:bcsz1}), we find
\begin{equation}
    H_1 = \frac{A_3}{2}\left(r^2 - \frac{1}{4}\right).
\end{equation} 

Having determined $H_i$ and hence $\psi_i$, we are now able to establish the boundary condition on the fluid potential $G$. At the base of the channel, (\ref{eq:bcspz0}) and (\ref{eq:bcspz1}) give $\nabla_{\perp} G= \boldsymbol{0}$ at $z = 0$ so that $G$ is a constant, which we set to be zero without loss of generality.
Integrating the following boundary conditions on string displacement at $z = 1$,
\begin{align}
     \partial_x G + \partial_y H_1 &= A_1 + A_3 y \label{eq:bcsz1transa} ,\\
    \partial_y G - \partial_x H_1 &= A_2 - A_3 x\label{eq:bcsz1transb},  
\end{align}
we find 
\begin{equation}
    G =  r \left(A_1 \cos \theta + A_2 \sin \theta \right) \quad \text{at} \quad z = 1,
\end{equation}
where without loss of generality, we have chosen zero constant of integration, fixing the final degree of freedom in $G.$

The boundary condition on the curved surface (\ref{eq:bcphiflux}) corresponding to no fluid flux through the outer wall is now
\begin{equation}
    (-\mathcal{K} \nabla G_{zz} + i \omega \nabla_{\perp_3} G ) \cdotp \boldsymbol{\hat{n} } = 0 \quad \text{on } \quad \partial \Omega_e. 
\end{equation}
For a cylindrical domain, we have
\begin{equation}
        -\kappa_1  G_{rzz} + i \omega G_r  = 0 \quad \text{on } \quad \partial \Omega_e. 
\end{equation}
The final boundary conditions come from imposing fluid pressure at each end of the domain (\ref{eq:bcphizzz}),
\begin{align}
    G_{zz} &= 0 \quad \text{at} \quad z = 0, \\
    G_{zz} &=0 \quad \text{at} \quad z = 1.
\end{align}

To summarise, it remains to solve the partial differential equation
\begin{equation}
     \nabla \cdotp \left( \mathcal{K} \nabla G_{zz} \right)  = \frac{i \omega}{\phi_f} \nabla^2_{\perp}G, \label{eq:gpde}
\end{equation}
subject to 
\begin{align}
    G = 0 \quad &\text{and} \quad  G_{zz} = 0 \quad \text{on} \quad z = 0, \label{eq:bcgz0}\\
    G = r(A_1 \cos \theta  + A_2 \sin \theta ) \quad &\text{and}\quad  G_{zz} = 0 \quad \text{at} \quad z = 1,\label{eq:bcgz1}\\
       -\kappa_1  G_{rzz} + i \omega G_r  = 0 \quad &\text{on} \quad \partial \Omega_e.\label{eq:bcnofouter}
\end{align}
Seeking a separable solution, we find that the boundary condition (\ref{eq:bcnofouter}) reduces to 
\begin{equation}
    G_r = 0 \quad \text{on} \quad \partial \Omega_e, \label{eq:bcgout}
\end{equation}
since the alternative $-\kappa_1 G_{zz} + i\omega G = 0 $ cannot be satisfied by (\ref{eq:gpde}) for real forcing frequencies. 
Imposing (\ref{eq:bcgz0}), (\ref{eq:bcgz1}) and (\ref{eq:bcgout}), we find the solution in cylindrical polar coordinates is 
\begin{equation}
    G =  \sum_{l = 1}^{\infty} J_1(\beta_l r) \left(b_l  \cos\theta + c_l \sin\theta \right) f_l(z),
\end{equation}
with coefficients 
\begin{equation}
    b_l = \frac{2 A_1 \beta_l J_2(\beta_l)}{ J_1^2(\beta_l)(\beta^2_l - 1)} \quad \text{and} \quad c_l = \frac{2 A_2 \beta_l J_2(\beta_l)}{ J_1^2(\beta_l)(\beta^2_l - 1)}\label{eq:expcoeff},
\end{equation}
where $J_n$ is a Bessel function and $f_{l}$ is the solution to the following ordinary differential equation with constant coefficients 
\begin{equation}
    \frac{d^4 f_{l}}{dz^4}  - \frac{\kappa_1 \beta^2_{l}}{\kappa_3} \frac{d^2 f
    _{l}}{dz^2} + \frac{i \beta_{l}^2 \omega }{\kappa_3 \phi_f}f_{l} = 0\label{eq:zode},
\end{equation}
normalised so that $f_{l} = 1$ at $z = 1$, i.e.
\begin{equation}
    f_l(z) = -\frac{\alpha_-^2}{\alpha_+^2 - \alpha_-^2}\frac{ \sinh{(\alpha_+ z)}}{{\sinh{(\alpha_+)}}} +\frac{\alpha_+^2}{\alpha_+^2 - \alpha_-^2}\frac{ \sinh{(\alpha_- z)}}{\sinh{(\alpha_-)}},\label{eq:fsol}
\end{equation}
where
\begin{align}
 \alpha_-^2 &= \frac{\beta_l^2 \kappa_1}{2 \kappa_3} \left(1 -  \sqrt{1 - \frac{4 i \omega \kappa_3}{\phi_f \beta_l^2 \kappa_1^2} }\right) \label{eq:alpham},\\
 \alpha_+^2 &= \frac{\beta_l^2 \kappa_1}{2 \kappa_3} \left(1 +  \sqrt{1 - \frac{4 i \omega \kappa_3}{ \phi_f \beta_l^2 \kappa_1^2} }\right).     \label{eq:alphap}
\end{align}
Here $\kappa_1$ and $\kappa_3$ are the components of the permeability tensor $\mathcal{K} = \rm{diag}(\kappa_1, \kappa_1, \kappa_3)$ and $\beta_{l}$ denote the zeros of $J'_{1}(\beta_{l}/2) = 0$. The oscillation frequency and the permeabilities appear only in the function $f_l$ - these parameters have a role in shaping the axial dependence of the solution, whereas the radial and azimuthal dependence is fixed by the spatial dependence of the boundary conditions imposed at the top and base of the domain. 

\begin{figure}
    \centering
    \includegraphics[width = \textwidth]{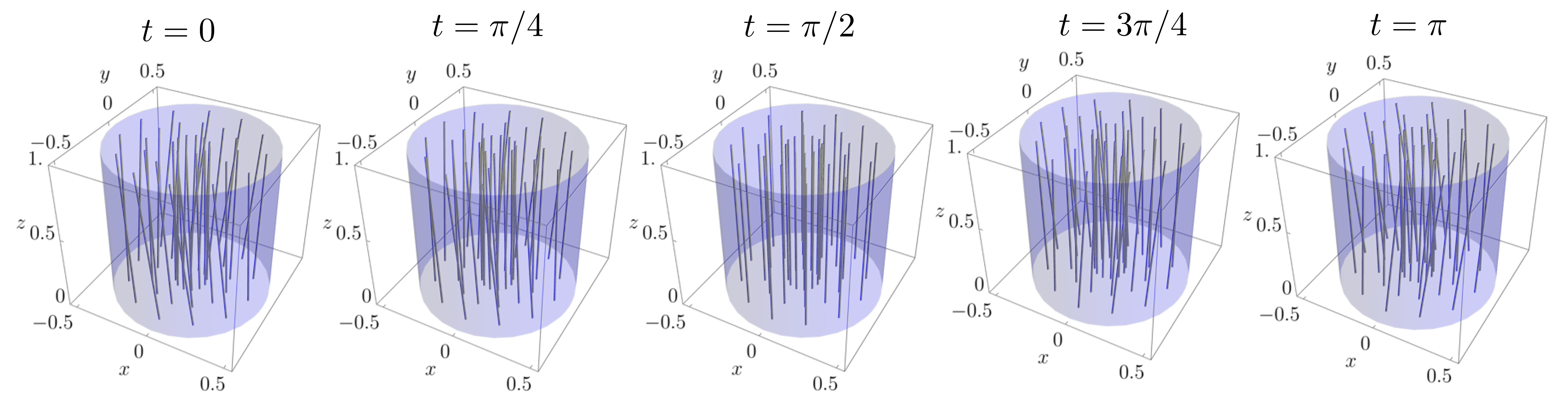}
    \caption{String displacement induced by twisting the upper plate. Plots use $\omega = 1$, $A_1 = 0$, $A_2 = 0 $ and $A_3 = 1$ and $\delta = 0.3$. The string profiles for an imposed are independent of the solid volume fraction.}
    \label{fig:twist}
\end{figure}

The corresponding displacement in a polar basis can be calculated from (\ref{eq:pots}) as
\begin{equation}
    \boldsymbol{s} = \Re \left[ \sum_{l}  \begin{pmatrix}
    \beta_l J_1'(\beta_l r) (b_l \cos \theta + c_l \sin \theta ) \\
   \frac{1}{r} J_1(\beta_l r) (c_l \cos \theta - b_l \sin \theta)
    \end{pmatrix} f_l(z) e^{i \omega t} \right] - A_3 r \cos{(\omega t)} z \boldsymbol{\hat{e}}_{\theta}.
\end{equation}
and fluid pressure is given by 
\begin{equation}
    p = \Re\left[ \sum_{l} J_1(\beta_l r) \left( b_l \cos \theta + c_l \sin \theta  \right) f_l^{''}(z)e^{i \omega t}\right].
\end{equation}
We now consider solutions corresponding to various choices of $A_1, A_2$ and $A_3$.
Imposing a twist on the upper plate
\begin{equation}
    A_1 = 0, \quad A_2 = 0, \quad A_3 \neq 0,
\end{equation}
we obtain the trivial solution for the fluid pressure corresponding to $p = 0$, and straight string profiles
\begin{equation}
    \boldsymbol{s} = - A_3 r z \cos(\omega t) \boldsymbol{\hat{e}}_{\theta},
\end{equation}
as shown in figure \ref{fig:twist}.
The absence of macroscale pressure gradients means there is no fluid-induced shear stress exerted at the string boundary - the fluid velocity equals the string velocity. 

The assumption of local periodicity at the microscale means that we can no longer satisfy a no-slip condition on the outer wall. This break down of periodicity at the outer wall creates a region of $O(\delta)$ where the homogenised model interfaces with flow governed by the Stokes equations. This forms a boundary layer where we would expect pressure gradients to exist, generating a microscale shear stress varying between adjacent strings. 

\begin{figure}
    \centering
    \includegraphics[width = \textwidth]{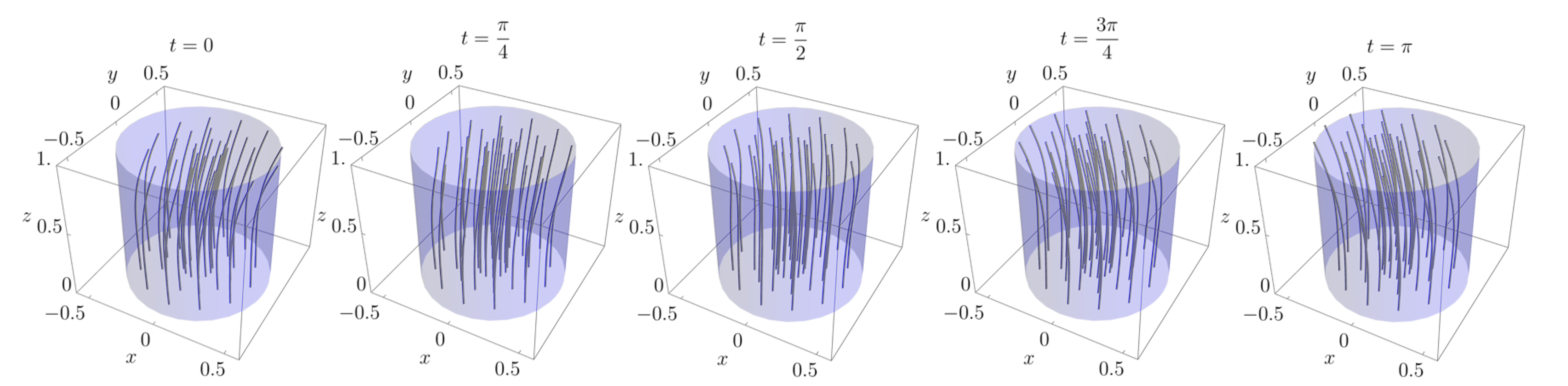}
    \caption{String displacement induced by translating the upper plate. Plots use $\omega = 1$, $A_1 = 1$, $A_2 = 0$, $A_3 = 0 $ and  $\delta = 0.1$ and $\phi_s = 0.2$ , corresponding to $\kappa_1 = 2.48 \times 10^{-3}$ and $\kappa_3 =4.18 \times 10^{-3}$.}
    \label{fig:translation}
\end{figure}

Pressure gradients on the macroscale are established when the upper end of the strings is translated relative to the base, corresponding to
\begin{equation}
    A_1 \neq 0, \quad A_2 = 0, \quad A_3 = 0 . 
\end{equation}
 These pressure gradients drive fluid flow which modifies the string profiles from the straight lines that would be observed in the absence of fluid (figure \ref{fig:translation}).  

The macroscale pressure gradients modulate the magnitude of the microscale shear stress. At the upper edge of the domain, we have a discontinuity in the radial derivatives of the fluid potential resulting from the scraping flow as the upper plate is translated relative to the outer domain wall, from boundary conditions (\ref{eq:bcgz1}) and (\ref{eq:bcgout}). Rescaling $r, z$ to obtain a problem in local coordinates, we solve to find the form of the singularity near $r = 1/2, z = 1$
\begin{equation}
    G_s \sim 2 A_1 \left(\frac{1}{2} - r \right)\ln{\left(\frac{1}{\kappa_{1}}(\frac{1}{2} - r)^2 + \frac{1}{\kappa_3} \left(1 - z \right)^2 \right)}  \cos \theta.\label{eq:sing}
\end{equation}
This behaviour near the edge of the domain means that the shear stress calculations from the homogenised model break down. While we cannot calculate the shear stress over the whole cross-section without including the effects from the boundary, we do gain insight into the axial variations through this example. The axial variation is set by the nondimensional frequency which describes the relative magnitude of the fluid stress and tension. In the next section, we quantify this effect by considering an example with regularised boundary conditions, chosen so that there is no discontinuity in the potential derivatives at the upper corner.

\subsubsection{Radial String Motion}\label{sec:radial}
We now suppose that the motion of the upper end of each string is no longer constrained to rigid-body motion, considering instead imposed displacements of the form
\begin{equation}
    \boldsymbol{s} =  (r - \frac{1}{2}) \cos(\omega t) \boldsymbol{\hat{e}}_r \quad \text{at} \quad z = 1\label{eq:bcradsz1},
\end{equation}
so that the strings do not pass through the boundary $r = 1/2.$ The strings are fixed at the base,
\begin{equation}
    \boldsymbol{s} = \boldsymbol{0} \quad \text{at} \quad z = 0. \label{eq:bcradsz0}
\end{equation}
As in section \ref{sec:platemotion}, there is no vertical pressure drop across the domain
\begin{equation}
    \varphi_{zz} = 0 \quad \text{at} \quad  z = 0, 1 ,
\end{equation}
and no fluid flows through the outer wall 
\begin{equation}
    \boldsymbol{u} \cdotp \boldsymbol{\hat{n}} = 0 \quad \text{at} \quad r = 1/2.
\end{equation}
Following a similar process to section \ref{sec:platemotion}, we seek time harmonic solutions (\ref{eq:thsol}). After imposing the boundary conditions on string displacement (\ref{eq:bcradsz1})-(\ref{eq:bcradsz0}), we find
\begin{equation}
    \psi_1 = 0 \quad \text{and} \quad 
    \psi_2 = 0 .
\end{equation}
The problem for the fluid potential (\ref{eq:macrog}) has boundary conditions
\begin{align}
    G =   \frac{r(r-1)}{2} \quad &\text{and} \quad G_{zz} = 0 \quad \text{at} \quad z = 1, \\
    G = 0 \quad &\text{and} \quad G_{zz} = 0 \quad \text{at} \quad z = 0,\\
    G_r = 0 \quad &\text{at} \quad \text{on} \quad \partial \Omega_e.
\end{align}
We find the solution using separation of variables
\begin{equation}
    G = d_0 + \sum_l d_l J_0 (\beta_l r) f_l (z),
\end{equation}
where $\beta_l$ satisfy $J'_0(\beta_l/2) = 0$, ordered so that $\beta_1 < \beta_2 < ...$ and $f_l$ is given by (\ref{eq:fsol}). The expansion coefficients are
\begin{equation}
    d_0 = -\frac{ 1}{48} \quad \text{and} \quad d_l = \frac{ 1 }{2}\frac{\int_{0}^{1/2} (r^3 - r^2)J_0(\beta_l r) dr}{ \int_0^{1/2} r J_0(\beta_l r) dr}.
\end{equation} 
The corresponding fluid pressure is
\begin{equation}
    p = \Re \left[ \sum_l d_l J_0(\beta_l r) f_l^{''}(z) e^{i\omega t}\right]\label{eq:pressuresolrad},
\end{equation}
and displacement is
\begin{equation}
    \boldsymbol{s} = \Re \left[ \sum_l d_l J_0'(\beta_l r) f_l(z)e^{i\omega t}\label{eq:dispsolrad} \right].
\end{equation}
Combining the macroscale pressure gradients with the microscale solution calculated in section \ref{sec:cellsolutions}, we can calculate the variation in microscale shear stress (\ref{eq:shear}) across the domain. 

\begin{figure}
    \centering
    \includegraphics[width = \textwidth]{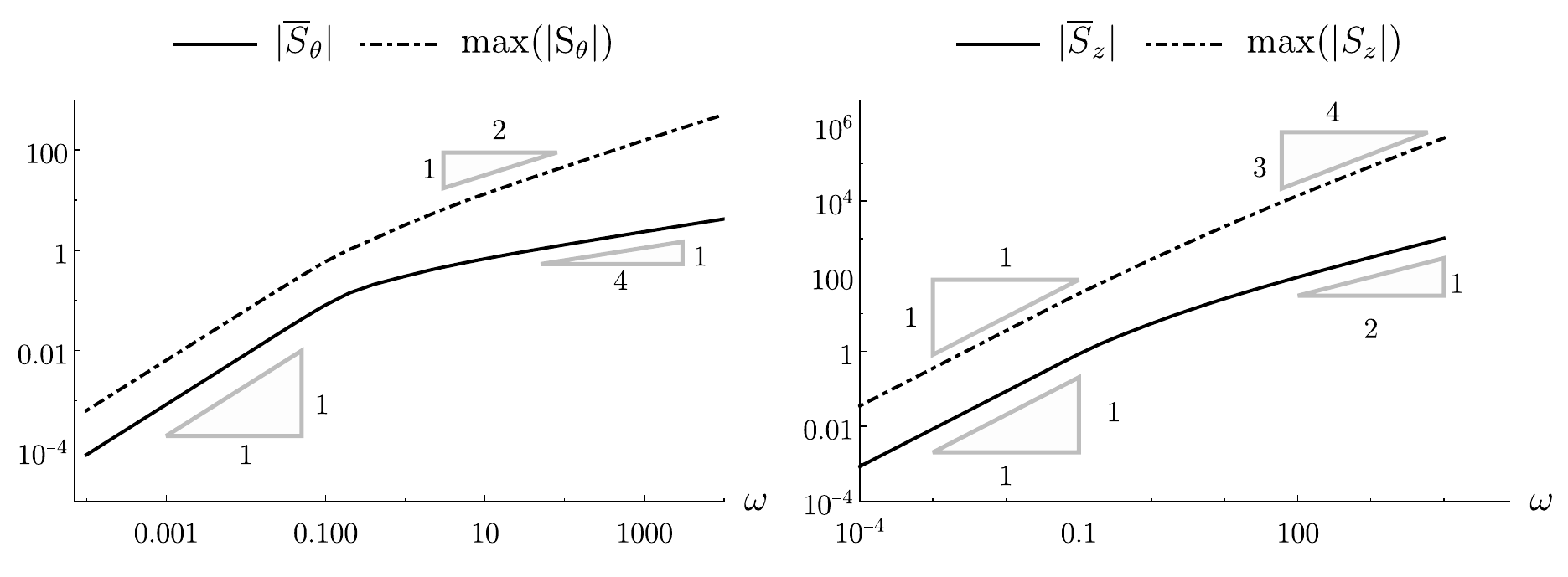}
    \caption{Frequency dependence of shear stress. Magnitudes of both the axial and transverse shear stress were evaluated at different points in space and time, using values of $M_z$ and $M_\theta$ for solid fraction $\phi_s = 0.2$ averaged around the string boundary. The maximum shear stress and averaged value are shown.  }
    \label{fig:sscalings}
\end{figure}

For a given imposed displacement, the magnitude of the shear stress, averaged over an oscillation cycle and spatially, first around a given string then over the domain, is dependent upon the dimensionless frequency $\omega$. We write the spatio-temporal average as 
\begin{equation}
    \overline{S} = \frac{1}{2\pi }\frac{1}{|\Omega|}\frac{1}{\tau}\int_{\Omega} \int_{0}^{\tau}\int_{\partial D_{s_0}} S(x, y, z, X, Y , t) dl_X dt dv.
\end{equation}
We can characterise the behaviour of this average in the limits of low and high frequency by expanding $(\ref{eq:fsol})$. For low forcing frequencies, we find the expansion of the second axial derivative of (\ref{eq:fsol}) becomes
\begin{equation}
    f_l''(z) \sim  \frac{i \omega}{\kappa_1 \phi_f} \left( z - \frac{\sinh(\sqrt{\frac{\kappa_1 \beta_l^2 }{\kappa_3}}z)}{\sinh(\sqrt{\frac{\kappa_1 \beta_l^2}{\kappa_3}})} \right) + O(\omega^2) \quad \text{as}\quad \omega \to 0,
\end{equation}
giving a linear dependence on the spatio-temporal average of the magnitude of the transverse and axial shear stress, calculated using (\ref{eq:shear}), as shown in figure \ref{fig:sscalings}.
The maximum values of the axial and transverse shear also scales linearly with the forcing frequency. Expanding for high frequency, we find
\begin{equation}
    \Re[f_l e^{i \omega t}] \sim \frac{1}{2} e^{a \lambda (z-1)} \cos(b \lambda (z - 1) -  \omega t)  + \frac{1}{2}e^{b\lambda (z-1)} \cos(a \lambda (z-1) + \omega t) \quad \text{as} \quad \omega \to \infty,
\end{equation}
where $a = \sin(\pi/8)$,  $b = \cos(\pi/8)$ and $\lambda = (\beta^2_l \omega /\phi_f \kappa_3)^{1/4}$. Thus, we have a boundary layer near the upper edge of the domain with a thickness $O(1/\lambda)$. Increasing the frequency of the imposed displacement decreases the width of the boundary layer, resulting in the average magnitudes of shear stress scaling less strongly than the maximum as shown in figure \ref{fig:sscalings}. This effect occurs as the dominant contributions to the averaged shear are confined to a region of $O(1/\lambda)$ near the upper edge.

\section{Discussion}\label{sec:discussion}
We have presented the derivation of a coarse-grained set of equations describing the fluid-structure interaction of many aligned strings and slow, viscous flow using homogenisation by multiscale asymptotics. We obtain three microscale problems that are independent of the slow-scale: two characteristic of the homogenisation of the incompressible Stokes equations in anisotropic porous media, and a third describing the role of string drag in inducing fluid flow. These problems capture the influence of string shape and solid fraction on the permeability coefficient which appears in the macroscale equations. The macroscale equations take the form of a poroelastic model for incompressible constituents, with a homogenised force balance on the strings taking the place of the Navier equation. To obtain the form of the homogenised force balance, we combined extensions to standard mulitple scales used to treat integral constraints in a slowly varying domain. For flow driven by an imposed time-harmonic motion of strings at the top of the domain, we have determined analytical solutions for a cylindrical domain. By combining the solutions for the macroscale pressure gradients with solutions to the cell problems, we are able to compute the spatial variation in microscale shear stress across the domain. 

For flow driven by imposed displacement of the strings, we found that the axial shear stress increased more rapidly with increasing frequency when compared with the transverse shear stress. In the high forcing frequency regime, the presence of a boundary layer at the upper edge of the domain resulted in the averaged shear stress scaling more slowly with oscillation frequency than the maximum. 

We considered fluid flow driven by the oscillation of the string ends, setting the fluid velocity scale via the string tension. The system can also be driven by an imposed flux or pressure drop, with results not presented here. The model also has analytical solutions for pressure gradients imposed across the domain, allowing the pressure drop to set the fluid velocity scale. Imposing an axial pressure drop and no string motion, we find a linear axial pressure gradient results, generating a uniform axial shear stress. Alternatively, the system can be driven by imposing a fluid flux at the boundary. If no string motion is imposed and a radially varying flux at the base drives fluid flow, we find the ratio of the transverse to axial permeability components is the parameter which determines the distribution of shear stress across the domain. Thanks to the linearity of the governing equations, we can combine the pressure- or flux-driven flow with our string-driven solution by superposition, allowing the combined influence on flow to be explored. 

The model developed here provides details of the derivation of a coarse-grained model of fluid-structure interaction between slow flow and assemblies of fibres using multiscale asymptotics. While we consider dense bundles of strings where a modified Darcy's law is applicable, this analysis may be extended lower solid fractions. Considering different dominant balances of terms in the Stokes momentum equation would result in a modified Brinkman equation when drag, viscous effects and pressure gradients balance, or Stokes flow when viscous terms balance pressure gradients \cite{Levy1983}. Extensions could incorporate other solid models to develop coarse-grained descriptions of fluid-structure interaction in fibrous beds where bending stiffness may be important. By using similar techniques to those used to model homogenisation of the bidomain equations for a beating heart \cite{Richardson2011}, the method could accommodate media where the solid fraction is slowly varying across the domain, including for example weak clumping effects.

The analytical solutions presented here provide insight into the dependence of fluid-string interaction on the solid fraction and driving frequency. The methodology allows for the incorporation of various microscale features and systematic identification of their effect on macroscale behaviour. The approach should complement other coarse-graining approaches and numerical techniques currently being used to study fluid-structure interaction at low Reynolds number.

\vspace{1em}
\textbf{Acknowledgements}
AK thanks Mohit Dalwadi for useful discussions and acknowledges support from the BBSRC under grant BB/M011224/1.

\bibliographystyle{siamplain}
\bibliography{2022-fluid-string-interaction-10.bib}

\begin{thebibliography}{10}

\bibitem{Alvarado2017}
{\sc J.~Alvarado, J.~Comtet, E.~Langre, and A.~E. Hosoi}, {\em {Nonlinear flow
  response of soft hair beds}}, Nature Physics, 13 (2017), pp.~1014--1019,
  \url{https://doi.org/10.1038/nphys4225}.

\bibitem{Artini2017}
{\sc G.~Artini and D.~Broc}, {\em A homogenisation method for a fsi problem:
  Application to a tube bundle row}, Pressure Vessels and Piping Conference, 5:
  High-Pressure Technology; ASME Nondestructive Evaluation, Diagnosis and
  Prognosis Division (NDPD); SPC Track for Senate (2017),
  \url{https://doi.org/10.1115/PVP2017-65737}.

\bibitem{Artini2018}
{\sc G.~Artini and D.~Broc}, {\em {Fluid Structure Interaction Homogenization
  for Tube Bundles: Significant Dissipative Effects}}, Pressure Vessels and
  Piping Conference, 4: Fluid-Structure Interaction (2018),
  \url{https://doi.org/10.1115/PVP2018-84344}.

\bibitem{Barry2001}
{\sc S.~I. Barry and M.~Holmes}, {\em {Asymptotic behaviour of thin poroelastic
  layers }}, IMA Journal of Applied Mathematics, 66 (2001), pp.~175--194,
  \url{https://doi.org/10.1093/imamat/66.2.175}.

\bibitem{Bennett2018}
{\sc T.~P. Bennett, G.~D'Alessandro, and K.~R. Daly}, {\em Multiscale models of
  metallic particles in nematic liquid crystals}, SIAM Journal on Applied
  Mathematics, 78 (2018), pp.~1228--1255,
  \url{https://doi.org/10.1137/18M1163919}.

\bibitem{Brinkman1949}
{\sc H.~C. Brinkman}, {\em A calculation of the viscous force exerted by a
  flowing fluid on a dense swarm of particles.}, applied Scientific Research, 1
  (1949), p.~341–360, \url{https://doi.org/10.1007/BF02120313}.

\bibitem{Bruna2015}
{\sc M.~Bruna and S.~J. Chapman}, {\em {Diffusion in spatially varying porous
  media}}, SIAM Journal on Applied Mathematics, 75 (2015), pp.~1648--1664,
  \url{https://doi.org/10.1137/141001834}.

\bibitem{Burridge1981}
{\sc R.~Burridge and J.~B. Keller}, {\em Poroelasticity equations derived from
  microstructure}, The Journal of the Acoustical Society of America, 70 (1981),
  pp.~1140--1146, \url{https://doi.org/10.1121/1.386945}.

\bibitem{us2022}
{\sc S.~J. Chapman, A.~Kent, J.~Oliver, and S.~Waters}, {\em {Integral
  Constraints in Multiple Scales Problems with a Slowly Varying Domain }}, In
  preparation.

\bibitem{Chapman2015}
{\sc S.~J. Chapman and S.~E. McBurnie}, {\em Integral constraints in
  multiple-scales problems}, European Journal of Applied Mathematics, 26
  (2015), p.~595–614, \url{https://doi.org/10.1017/S0956792514000412}.

\bibitem{Dalwadi2015}
{\sc M.~P. Dalwadi, I.~M. Griffiths, and M.~Bruna}, {\em {Understanding how
  porosity gradients can make a better filter using homogenization theory}},
  Proceedings of the Royal Society A: Mathematical, Physical and Engineering
  Sciences, 471 (2015), \url{https://doi.org/10.1098/rspa.2015.0464},
  \url{https://arxiv.org/abs/1507.01722}.

\bibitem{Drew1983}
{\sc D.~A. Drew}, {\em Mathematical modeling of two-phase flow}, Annual Review
  of Fluid Mechanics, 15 (1983), pp.~261--291,
  \url{https://doi.org/10.1146/annurev.fl.15.010183.001401}.

\bibitem{Drummond1984}
{\sc J.~Drummond and M.~Tahir}, {\em Laminar viscous flow through regular
  arrays of parallel solid cylinders}, International Journal of Multiphase
  Flow, 10 (1984), pp.~515--540,
  \url{https://doi.org/https://doi.org/10.1016/0301-9322(84)90079-X}.

\bibitem{DuRoure2019}
{\sc O.~du~Roure, A.~Lindner, E.~N. Nazockdast, and M.~J. Shelley}, {\em
  {Dynamics of Flexible Fibers in Viscous Flows and Fluids}}, Annual Review of
  Fluid Mechanics, 51 (2019), pp.~539--572,
  \url{https://doi.org/10.1146/annurev-fluid-122316-045153}.

\bibitem{Duprat2016}
{\sc C.~Duprat and H.~Stone}, eds., {\em Fluid-Structure Interactions in
  Low-Reynolds-Number Flows}, Soft Matter Series, The Royal Society of
  Chemistry, 2016, \url{https://doi.org/10.1039/9781782628491}.

\bibitem{Gopinath2011}
{\sc A.~Gopinath and L.~Mahadevan}, {\em Elastohydrodynamics of wet bristles,
  carpets and brushes}, Proceedings of the Royal Society A: Mathematical,
  Physical and Engineering Sciences, 467 (2011), pp.~1665--1685,
  \url{https://doi.org/10.1098/rspa.2010.0228}.

\bibitem{Happel1959}
{\sc J.~Happel}, {\em Viscous flow relative to arrays of cylinders}, AIChE
  Journal, 5 (1959), pp.~174--177,
  \url{https://doi.org/https://doi.org/10.1002/aic.690050211}.

\bibitem{Hinch1991}
{\sc E.~J. Hinch}, {\em Perturbation Methods}, Cambridge Texts in Applied
  Mathematics, Cambridge University Press, 1991,
  \url{https://doi.org/10.1017/CBO9781139172189}.

\bibitem{Holmes2013}
{\sc M.~H. Holmes}, {\em Introduction to Perturbation Methods}, vol.~20 of
  Texts in Applied Mathematics, Springer, New York, NY, 2~ed., 2013.

\bibitem{Hosoi2019}
{\sc A.~E. Hosoi}, {\em Corrsin lecture on hairy hydrodynamics}, Physical
  Review Fluids, 4 (2019), p.~110508,
  \url{https://doi.org/10.1103/PhysRevFluids.4.110508}.

\bibitem{Keller1980}
{\sc J.~B. Keller}, {\em Darcy’s law for flow in porous media and the
  two-space method}, in Nonlinear partial differential equations in engineering
  and applied science, Routledge, 1980, pp.~429--443.

\bibitem{Lauga2016}
{\sc E.~Lauga, C.~J. Pipe, and B.~Le~Révérend}, {\em Sensing in the mouth: A
  model for filiform papillae as strain amplifiers}, Frontiers in Physics, 4
  (2016), p.~35, \url{https://doi.org/10.3389/fphy.2016.00035}.

\bibitem{Levy1983}
{\sc T.~Lévy}, {\em Fluid flow through an array of fixed particles},
  International Journal of Engineering Science, 21 (1983), pp.~11--23,
  \url{https://doi.org/https://doi.org/10.1016/0020-7225(83)90035-6}.

\bibitem{Maganaris2017}
{\sc C.~N. Maganaris, P.~Chatzistergos, N.~D. Reeves, and M.~V. Narici}, {\em
  {Quantification of internal stress-strain fields in human tendon: Unraveling
  the mechanisms that underlie regional tendon adaptations and mal-adaptations
  to mechanical loading and the effectiveness of therapeutic eccentric
  exercise}}, Frontiers in Physiology, 8 (2017), pp.~1--11,
  \url{https://doi.org/10.3389/fphys.2017.00091}.

\bibitem{Mouthuy2022}
{\sc P.~A. Mouthuy, S.~Snelling, R.~Hostettler, A.~Kharchenko, S.~Salmon,
  A.~Wainman, J.~Mimpen, C.~Paul, and A.~Carr}, {\em Humanoid robots to
  mechanically stress human cells grown in soft bioreactor}, submitted.

\bibitem{Nasto2018}
{\sc A.~Nasto, P.-T. Brun, and A.~E. Hosoi}, {\em Viscous entrainment on hairy
  surfaces}, Physical Review Fluids, 3 (2018), p.~024002,
  \url{https://doi.org/10.1103/PhysRevFluids.3.024002}.

\bibitem{Nasto2016}
{\sc A.~Nasto, M.~Regli, P.-T. Brun, J.~Alvarado, C.~Clanet, and A.~E. Hosoi},
  {\em Air entrainment in hairy surfaces}, Phys. Rev. Fluids, 1 (2016),
  p.~033905, \url{https://doi.org/10.1103/PhysRevFluids.1.033905}.

\bibitem{Nazockdast2017}
{\sc E.~Nazockdast, A.~Rahimian, D.~Zorin, and M.~Shelley}, {\em A fast
  platform for simulating semi-flexible fiber suspensions applied to cell
  mechanics}, Journal of Computational Physics, 329 (2017), pp.~173--209,
  \url{https://doi.org/https://doi.org/10.1016/j.jcp.2016.10.026}.

\bibitem{Ockendon1993}
{\sc H.~Ockendon and E.~L. Terrill}, {\em A mathematical model for the
  wet-spinning process}, European Journal of Applied Mathematics, 4 (1993),
  p.~341–360, \url{https://doi.org/10.1017/S0956792500001170}.

\bibitem{Richardson2011}
{\sc G.~Richardson and S.~J. Chapman}, {\em {Derivation of the bidomain
  equations for a beating heart with a general microstructure}}, SIAM Journal
  on Applied Mathematics, 71 (2011), pp.~657--675,
  \url{https://doi.org/10.1137/090777165}.

\bibitem{Rooney2021}
{\sc C.~M. Rooney, C.~P. Please, and S.~D. Howison}, {\em Homogenisation
  applied to thermal radiation in porous media}, European Journal of Applied
  Mathematics, 32 (2021), p.~784–805,
  \url{https://doi.org/10.1017/S0956792520000388}.

\bibitem{AlnaesBlechta2015}
{\sc A.~M. S., J.~Blechta, J.~Hake, A.~Johansson, K.~B., A.~Logg, R.~C.,
  J.~Ring, M.~E. Rognes, and G.~N. Wells}, {\em The fenics project version
  1.5}, Archive of Numerical Software, 3 (2015),
  \url{https://doi.org/10.11588/ans.2015.100.20553}.

\bibitem{Sparrow1959}
{\sc E.~M. Sparrow and A.~L. Loeffler~JR.}, {\em Longitudinal laminar flow
  between cylinders arranged in regular array}, AIChE Journal, 5 (1959),
  pp.~325--330, \url{https://doi.org/https://doi.org/10.1002/aic.690050315}.

\bibitem{Stein2019}
{\sc D.~B. Stein and M.~J. Shelley}, {\em Coarse graining the dynamics of
  immersed and driven fiber assemblies}, Physical Review Fluids, 4 (2019),
  p.~073302, \url{https://doi.org/10.1103/PhysRevFluids.4.073302}.

\bibitem{Szaniawski1977}
{\sc A.~Szaniawski}, {\em Equations of steady flow through slightly curved
  multifilament bundles.}, Archives of Mechanics, 29 (1977), pp.~519--530.

\bibitem{Terrill1994}
{\sc E.~L. Terrill and J.~G. Byatt-Smith}, {\em A mathematical model for
  washing a tow of fibres: Part 1}, European Journal of Applied Mathematics, 5
  (1994), p.~525–535, \url{https://doi.org/10.1017/S0956792500001595}.

\bibitem{Thomazo2020}
{\sc J.~Thomazo, E.~Lauga, B.~Le~R\'ev\'erend, E.~Wandersman, and A.~M.
  Prevost}, {\em Collective stiffening of soft hair assemblies}, Phys. Rev. E,
  102 (2020), p.~010602, \url{https://doi.org/10.1103/PhysRevE.102.010602}.

\end{thebibliography}

\end{document}